\documentclass[letterpaper,twocolumn,10pt]{article}
\usepackage{usenix}

\usepackage{etoolbox}

\usepackage{tikz}
\usepackage{amsmath}

\usepackage{tikz}
\usetikzlibrary{decorations}
\usetikzlibrary{arrows}
\usetikzlibrary{shapes}
\usetikzlibrary{math}

\usepackage{amsfonts}
\usepackage{cleveref}
\usepackage{booktabs}
\usepackage{multirow}
\usepackage{algorithm}
\usepackage{algpseudocode}
\usepackage{xspace}

\usepackage[normalem]{ulem}
\useunder{\uline}{\ul}{}

\newcommand{\tzfontsize}[0]{\footnotesize}

\newcommand{\sys}{\textsc{FastDecode}}
\newcommand{\thetitle}{\sys{}: High-Throughput GPU-Efficient LLM Serving using Heterogeneous Pipelines}
\newcommand{\kvcache}{\textit{KV-cache}}
\newcommand{\stype}{\rpart{}}
\newcommand{\btype}{\spart{}}
\newcommand{\rpart}{\textit{R-Part}}
\newcommand{\spart}{\textit{S-Part}}

\newcommand{\speedupu}{$1.88\times - 5.04\times$}

\newcommand{\noneed}[1]{}

\definecolor{colorA}{RGB}{244,177,131} 
\definecolor{colorB}{RGB}{255,217,102} 
\definecolor{colorC}{RGB}{197,224,180} 
\definecolor{colorD}{RGB}{189,215,238} 
\definecolor{colorE}{RGB}{141,211,199} 
\definecolor{colorF}{RGB}{190,186,218} 
\definecolor{colorG}{RGB}{251,128,114} 
\definecolor{colorH}{RGB}{128,177,211} 
\definecolor{colorI}{RGB}{253,180,98} 
\definecolor{colorJ}{RGB}{204,204,204} 
\definecolor{colorK}{RGB}{252,205,229} 
\definecolor{colorL}{RGB}{179,222,105} 
\definecolor{colorM}{RGB}{255,217,47} 
\definecolor{colorN}{RGB}{252,141,89} 
\definecolor{colorO}{RGB}{116,169,207} 
\definecolor{colorP}{RGB}{102,194,164} 
\definecolor{colorQ}{RGB}{244,161,67} 
\definecolor{colorR}{RGB}{255,201,54} 
\definecolor{colorS}{RGB}{120,198,121} 

\newcommand{\colorinter}[0]{colorM}
\newcommand{\colorkv}[0]{colorI}
\newcommand{\colormodel}[0]{colorC}
\newcommand{\colorsys}[0]{colorE}
\newcommand{\colorba}[0]{colorD}
\newcommand{\colorbb}[0]{colorK}

\newcommand{\colorcpu}[0]{colorO}
\newcommand{\colorgpu}[0]{colorL}

\begin{document}

\date{}

\title{\thetitle{}}

\author{Jiaao He \\ Tsinghua University
\and Jidong Zhai \\ Tsinghua University}

\maketitle

\begin{abstract}
Cost of serving large language models (LLM) is high,
but the expensive and scarce GPUs are poorly efficient when generating tokens sequentially, unless the batch of sequences is enlarged.
However, the batch size is limited by some constantly reused intermediate results, namely KV-Cache.
They occupy too much memory to fit more sequences into a GPU simultaneously.
While they could be offloaded to host memory, the CPU-GPU bandwidth is an inevitable bottleneck.

We find a way to decompose the transformer models into two parts of different characteristics, one of which includes the memory-bound KV-Cache accessing.
Our key insight is that the aggregated memory capacity, bandwidth, and computing power of CPUs across multiple nodes is an efficient option to process this part.
Performance improvement comes from reduced data transmission overhead and boosted GPU throughput to process the other model part.
Moreover, we address efficiency challenges brought by heterogeneity at both temporal and inter-device scopes using scheduling and performance modeling techniques.
Evaluation results show that our system achieves \speedupu{} the throughput of vLLM when serving modern LLMs with the same GPU.
\end{abstract}

\section{Introduction}

The large language models (LLM) are gaining high attention.
These transformer-based models are very hardware-friendly when training and evaluating~\cite{megatron,bagualu},
because the main computation workload is matrix multiplication, a highly optimized operation to run on accelerators, e.g., GPUs.
However, when using the models, the auto-regressive procedure, i.e., decoding, is inefficient. 
Because tokens in a sequence are generated one-by-one, one of the operand matrices is in fact a vector.
Multiplying a vector with a matrix achieves much lower throughput due to poor utilization of GPUs.

Although employing an arbitrary number of GPUs can always increase the throughput, in most cases, GPU is one of the most scarce resources.
So, we should increase the utilization of single GPUs for many reasons, including affordability and being eco-friendly.

Enlarging batch size, i.e., generating tokens for multiple requests simultaneously, is the most feasible way without changing the model.
Different from prior use cases of neural networks, there is a larger opportunity for LLMs to run generation in batches, as an LLM usually serves many users online. 
The latency requirement to generate a token is much looser than other user cases of NN models, e.g., object detection in autonomous driving.
Keeping up with reading speed of the user is the most strict latency requirement for text generation.

\begin{figure}[ht]
    \centering
    \includegraphics[scale=.5]{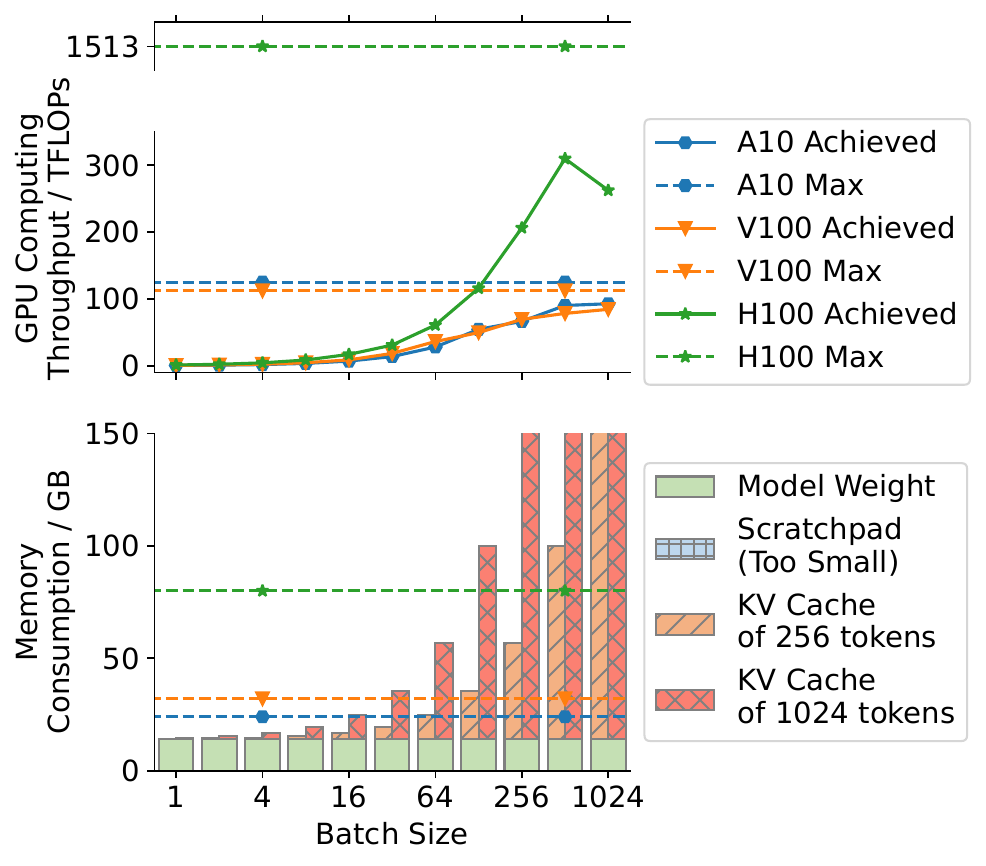}
    \flushleft
    \caption{Memory footprint of \kvcache{} stops increasing GPU utilization by enlarging batch size}
    \label{fig:perfmot}
\end{figure}

However, generating a new token depends on huge intermediate results of generating the previous tokens, namely \kvcache{}~\cite{kvcache}.
Processing batched requests results in a much larger memory footprint, far beyond the capacity of GPU memory.
\Cref{fig:perfmot} shows the dilemma instantiated on a common 7b model and several different GPUs.
Increasing batch size makes the GPUs significantly better utilized, but the memory footprint of the \kvcache{} is much larger than the GPU memory.
To make it worse, the \kvcache{} becomes even larger as more tokens are generated and the sequences get longer.

Host memory has naturally become the place to offload the KV-cache~\cite{vllm}, as it is larger and cheaper than GPU memory.
However, the \kvcache{} is not cold data.
The complete \kvcache{} is loaded into GPU memory to generate every token.
Considering the enormous size of \kvcache{} as suggested in \Cref{fig:perfmot},
transmitting it between GPU and host memory frequently is the bottleneck of the offloading design.
Essentially, the bandwidth of PCIe is always much lower than the memory bandwidth of GPUs and even CPUs.

\begin{figure}[ht]
    \centering
    \includegraphics[scale=.5]{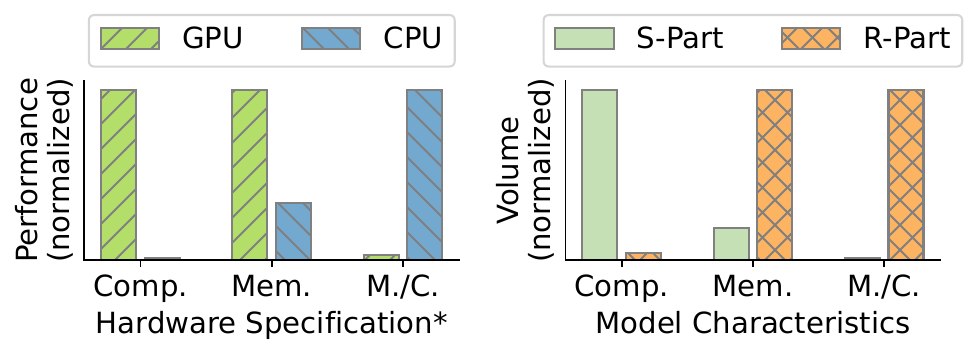}
    \flushleft
    {\footnotesize * Accurate numbers are in \Cref{sec:cpu}}
    \caption{Performance characteristics of typical GPUs and CPUs, matching the need of two parts of the model}
    \label{fig:perfhw}
\end{figure}

We study the performance characteristics, including compute throughput and memory bandwidth, of both GPUs and CPUs, as shown in \Cref{fig:perfhw}.
We find that compared with the huge gap in compute power, the two types of hardware have a much closer gap in memory bandwidth.

Fortunately, we find a way to partition the transformer model into two parts, namely \rpart{} and \spart{}.
\kvcache{} is included by the former.
Little performance loss is introduced when completely moving the memory-bound part to the host side,
as the ratio of memory bandwidth and compute throughput can fulfill its requirement.
Therefore, we get \textbf{our key insight} that we should \textbf{compute near KV-cache} on CPUs.
Instead of transmitting KV-cache data over any inter-device connection, we transmit the activation tensors, which are orders of magnitudes smaller than KV-cache.

Our approach totally removes intermediate data of sequences, the \kvcache{}, from GPU memory.
Therefore, the batch size can be greatly increased, and the GPUs can be optimally utilized.
However, such a heterogeneous approach faces three challenges to achieve high overall throughput.

\newcommand{\cci}[2]{\textit{\textbf{#1 #2:}}}

\cci{Challenge}{1} The CPU is busy but slow.
It runs multiple tasks, including batch gathering, tokenization, and coordinating the GPUs.
Performing extra computation interferes with these tasks, slowing down all of them.
To add to the difficulty, the memory bandwidth of a CPU is lower than \noneed{that of a}GPU. 

\cci{Challenge}{2} The pattern of workload variation, as the generated sequences get longer, differs between the two parts.
In our solution, the CPU and the GPU take turns to perform computation, and pass the results to each other.
A basic pipeline of multiple batches of requests is used to utilize both of them.
However, computation on the CPU takes longer time as the generated sequence gets longer, while the latency of its counterpart on GPU does not change at all.
This makes it hard to always utilize both CPU and GPU.

\cci{Challenge}{3} Careful orchestration is needed to balance the performance of both types of hardware.
Bottleneck may be either of the GPU or CPU, because they are tightly coupled.
We need to balance the two considering the heterogeneous hardware and token generation workload.
We seek for a minimum CPU requirement that can fully exploit the compute power of the GPU, aiming at minimizing the overall cost.

Our system, \sys{}, is a CPU-GPU heterogeneous pipeline for LLM inference that addresses the challenges by the following innovations.

\cci{Innovation}{1} We employ multiple out-of-chassis remote CPUs for \kvcache{} and the related computation.
The aggregated memory capacity and bandwidth of the system are scaled up.
The distributed CPUs can achieve sufficient throughput to saturate the GPU, with moderate communication overhead.

\cci{Innovation}{2} We invent a sequence-level load-stabilizing schedule to minimize idling and better utilize both types of hardware.
The workload on a CPU is proportional to the total length of sequences it maintains.
To keep the latency stable, sequences are fed into the system following a workload control algorithm.
Short and long sequences are simultaneously processed by CPU workers, leaving the total length of sequences stable.
As a result, the overall latency of CPUs changes more gently, and both types of hardware are better utilized.

\cci{Innovation}{3} We adopt a model-guided approach to orchestrate the GPU with CPUs.
It quantitatively characterizes the performance bottleneck considering different aspects of the LLM inference tasks.
Aggregated memory bandwidth is identified as the key metric in selecting the CPUs.
For a given model and GPU setup, based on \noneed{the} profiling result of a micro-benchmark, we can estimate the minimum required aggregated CPU memory bandwidth for different batch sizes.

Overall, the throughput of a single GPU is saturated with a significantly larger batch size.
Thanks to the scalability and aggregated power of CPUs across nodes, high overall token generation throughput is achieved with affordable GPU resources.
In our evaluation, up to $5\times$ throughput of vLLM is achieved on the same GPU with acceptable latency.

Contributions of this paper are summarized as follows.

\begin{itemize}
    \item We find an unconventional way to decompose the auto-regressive transformer model, with high potential of performance improvement.
    \item We propose a near-memory processing system over the \kvcache{} that exploits the aggregated memory bandwidth out-of-chassis CPUs for higher throughput.
    \item We invent a sequence-level pipeline schedule to balance the growing-with-time workload and the fixed workload in token generation using LLMs.
    \item We create a performance model that can provide optimal hardware configuration for different models and requirements using our system.
\end{itemize}

This paper is organized as follows.
\Cref{sec:bkg} provides background information of LLMs and hardware options to serve them.
\Cref{sec:insight} shows our way to decompose the model, and illustrates our key insight that processing near \kvcache{} can boost overall throughput.
\Cref{sec:overview} introduces the design of our system, with techniques to resolve challenges brought by heterogeneity in both terms of workload and hardware.
\Cref{sec:impl} includes more details in our implementation.
\Cref{sec:eval} compares the performance of our system with other systems, and \Cref{sec:ablation} shows more experiment results for analyzing our performance.
\Cref{sec:related} includes discussion with more related works, and \Cref{sec:conclu} concludes our work.
\section{Background and Motivation}
\label{sec:bkg}

\subsection{Transformer Model and KV-Cache}

The auto-regressive models are based on transformer structure~\cite{attention}.
The key module of these models is the \textit{attention layer}.
We briefly illustrate its process as follows.
Denote the feature vector of the i-th token in a sequence as $X_i$.

First, $X_i$ is mapped to three different linear spaces, which is implemented by three fully-connected layers.

\begin{equation}
\begin{aligned}
    Q_i & = W_q X_i \\
    K_i & = W_k X_i \\
    V_i & = W_v X_i
\end{aligned}
\end{equation}

For the i-th token, inner production is applied between its feature vector and the $K_j$ feature vectors of all tokens before it. 
This is actually the attention process, and it generates an attention vector for the current token.

\begin{equation}
\label{equ:attn}
    A_{i} = \text{Normalize} \{ Q_i \cdot K_j (j=1, \dots, i-1)\}
\end{equation}

The attention vector is normalized, commonly using \textit{softmax} operation.
Then, it is used as a weight to gather information, i.e., add up the $V_j$ vectors of all preceding tokens.

\begin{equation}
\label{equ:weisum}
    O_{i} = \sum_{j=1}^{i-1} A_{ij} V_j
\end{equation}

The output is applied with the final linear transformation using another fully-connected layer.

\begin{equation}
\label{equ:out}
    Y_{i} = W_o O_i
\end{equation}

A \textit{transformer block} consists of one attention layer followed by a multi-layer perceptron (\textit{MLP}) module, 
which includes multiple large fully-connected layers and non-linear activation functions in between.
Connecting tens of such transformer blocks sequentially makes a complete decoder model, which is the backbone of most well-known LLMs, including GPT~\cite{gpt2,gpt4}, Llama~\cite{llama,llama2}, and many more.

When using such models in real-world tasks including chat and text generation, tokens are produced one-by-one.
To get the next token, only the latest token needs to be processed by the model, because the fully-connected layers and MLPs process each token independently.
However, \cref{equ:attn} and \cref{equ:weisum} of the attention layer involve reaction between the latest token and all previous tokens.
Instead of re-computing them, $K_j$ and $V_j$ can be saved in memory and reused for the newly generated tokens.
\kvcache{}~\cite{kvcache} refers to these saved intermediate tensors.
For a sequence of length $S$, this technique reduces the total amount of inner production computation between feature vectors from $O(S^3)$ to $O(S^2)$.
So, using \kvcache{} is mandatory in LLM inference.

\subsection{Accelerating Decoding}

There are two steps to use an LLM to respond to a request.
First, during the \textbf{prefilling} stage, the entire input sequence from the user is processed by the model, where all the tokens in the sequence can be processed as a batch in the MLP layers.
Then, in the \textbf{decoding} stage, the model use the feature vector of the last known token to predict the next token to be appended to the sequence.
So, each new token of the generated sequence goes through the model one-by-one.

Computation efficiency is extremely important, as it is directly related to the cost of serving LLMs.
Unfortunately, using GPUs, generating a single sequence is poorly efficient, because the main computation workload is applying fully connected layers to one feature vector in the decoding stage.
In other words, the main computation task is multiplying matrices with vectors (GeMV).
There is little chance to reuse the matrix data in near-processor memory, so accessing the global memory bounds the workload.
The numerous floating point number units on GPUs are underutilized.

To leverage the computation throughput of GPUs, enlarging batch size is the most feasible approach.
A batch of sequences are generated simultaneously, so multiple tokens are processed at the same time.
The feature vectors are stacked together to be a matrix, and the GeMV computation becomes multiplying the weight matrix with the feature matrix (GeMM). 
GeMM is a highly optimized operation on GPUs.
As long as the batch size is large enough, the computation power of GPUs is fully exploited.

Specifically, for the auto-regressive transformer-based models, Orca~\cite{orca} points out that the granularity of batching can be reduced to improve performance.
Instead of batching complete sequences together, it is more effective to batch the generation task of single tokens.
This technique greatly improves the throughput of serving LLMs by introducing more chances of batching.

Unfortunately, beside leaving the memory issue of \kvcache{} unaddressed, the flexible batching mechanism in Orca introduces significant memory fragmentation.
Paged attention technique is adapted by vLLM~\cite{vllm} to address the memory issue.
GPU and host memory of \kvcache{} is managed by pages, so that the GPU memory can be better utilized without fragmentation. 
Also, host memory can be used to store more \kvcache{} for more sequences, so the chance for batching token generation tasks is increased.

Chance of batching in vLLM is still limited, because swapping the large \kvcache{} over PCIe between GPU and host memory introduces high overhead.
Therefore, vLLM has to reduce the swapping frequency.
So, as the sequences get longer, \kvcache{} of few sequences can reside in the GPU memory, resulting in a small batch size.
The system achieves high throughput in less common cases, e.g., generating tokens with shared prefix or wide beam searching,
where multiple independent new tokens share the same \kvcache{}.

FlexGen~\cite{flexgen} studies on finding an optimal offloading order of both model weights and \kvcache{}.
Still, the \kvcache{} is orders of magnitudes larger than the model weights. 
Transmitting them over the PCIe link, which is much slower than the memory bandwidth,
is inherently inefficient.

In summary, it is a consensus that increasing batch size is the most effective way for higher throughput. 
However,  because \kvcache{} has to be in the GPU memory for computation,
few works can achieve actual speed up due to its large memory footprint.

Differently, we find that \kvcache{} does not need to present in GPU memory.
In this paper, we show challenges and our solutions that unleashes the power of CPUs to handle the \kvcache{} and achieve high token generating throughput by enabling a significantly larger batch size.

\subsection{Memory-bound Workload Fits CPU}
\label{sec:cpu}

While the slower but larger host memory is used to compensate for the lack of GPU memory capacity, the CPUs are barely used to perform computation.
They have up to TFLOPs of floating point computation throughput,
negligible compared with hundreds of TFLOPs achieved by specialized tensor processing units on GPUs.

\begin{table}[h]
    \caption{Performance and Power Comparison}
    \tzfontsize{}
    \centering 

\begin{tabular}{@{}cclllll@{}}
\toprule
\multirow{2}{*}{Type} & \multirow{2}{*}{Model} & \multicolumn{1}{c}{\multirow{2}{*}{TDP}} & \multicolumn{2}{c}{Compute}                             & \multicolumn{2}{c}{Memory}                             \\
                      &                        & \multicolumn{1}{c}{}                     & \multicolumn{1}{c}{FLOPs} & \multicolumn{1}{c}{W. per.} & \multicolumn{1}{c}{GB/s} & \multicolumn{1}{c}{W. per.} \\ \midrule
\multirow{2}{*}{CPU}  & Xeon*                  & 125 W                                    & 1.3 T                     & 96.15                       & 128                      & 0.97                        \\
                      & Epyc*                  & 155 W                                    & 1.2 T                     & 129.2                       & 205                      & 0.76                        \\
\multirow{2}{*}{GPU}  & A10                    & 150 W                                    & 125 T                     & 1.2                         & 600                      & 0.25                        \\
                      & V100                   & 250 W                                    & 112 T                     & 2.2                         & 900                      & 0.27                        \\ \bottomrule
\end{tabular}

\flushleft
{\footnotesize * Using Intel Xeon Gold 5218 and AMD Epyc 7452 CPUs.}
    \label{table:hwcost}
\end{table}

However, in terms of memory access bandwidth, the gap between CPU and GPU is smaller.
\Cref{table:hwcost} lists the compute throughput and memory bandwidth of several common ones.
Modern server-class CPUs can achieve hundreds of GB/s.
Memory bandwidth of mid-range GPUs, e.g. NVIDIA A10, is only a few times larger than the CPUs.
A dual-socket AMD Epyc server can achieve $68\%$ of its memory bandwidth.
Even the top-level GPUs can barely have more than $10\times$ bandwidth.
Additionally, different from the huge gap between GPUs of different levels,
the memory bandwidth remain similar from entry-level to high-end in each generation of CPUs.

CPUs are more attractive, considering the cost.
As a general purpose processor that exists in every computer, they are much more widely deployed than GPUs.
It is much easy to acquire a large number of CPUs with relatively low cost. 
We can easily enlarge the memory capacity and bandwidth by adding standard DIMMs to the servers.
On the contrary, GPU memory is not only expensive but also hard to extend its capacity as the memory is soldered on the circuit.

\Cref{table:hwcost} includes power consumption of the hardware as a metric of efficiency.
Maximum power consumption of CPUs is a few times lower than GPUs.
Besides, when memory access is the major workload, the hardware usually does not consume as much power as the TDP. 
So, the actual efficiency gap is even smaller than our estimation.

In conclusion, CPU is an appealing option for the memory-bound jobs.

\section{Observation and Insights}
\label{sec:insight}
\subsection{Performance Dilemma and Decomposition}

\begin{figure}[h]
    \centering
    \begin{tikzpicture}
\tzfontsize{}

\tikzmath{
    \itemh = 4;
    \itemw = 10;
    \paramx = 46;
    \paramy = 16;
    \kvx = 14;
    \yh = 16;
    \qparamy = \paramy + 8;
    \oparamy = \paramy - \yh - 8;
}

\node[fill=\colormodel!40!white,draw=gray,minimum width=\itemw mm, minimum height=14mm,align=center] (pa) at (\paramx mm, \qparamy mm) {Model\\ Param.\\ $W_{q,k,v}$};
\node[fill=\colormodel!60!white,draw=gray,minimum width=\itemw mm, minimum height=16mm ,align=center] (pb) at (\paramx mm, \oparamy mm) {Model\\ Param.\\ $W_o$ \& MLP};

\foreach \rowi in {0,1} {
    \tikzmath{
        \px = \paramx * (0.5 + \rowi);
        \py = \paramy + \yh;
        if \rowi==0 then {
            \bs = 1;
            let \bcolor = red;
        } else {
            \bs = 4;
            let \bcolor = green;
        };
    }
    \foreach \bi in {1,...,\bs} {
        \tikzmath{
            \bx = \px - \bi * 1;
            \by = \py + (\bs - \bi) * 1;
            \iy = \paramy + (\bs - \bi) * 1;
            \kx = \px + \kvx - \bi * 1; 
            \qy = \paramy - \yh + (\bs - \bi) * 1;
            \ky = \qy + \yh / 2;
            \oy = \paramy - \yh * 2 + (\bs - \bi) * 1;
        }
        \node[fill=gray!10!white,draw=gray,minimum width=\itemw mm, minimum height=\itemh mm] (x\rowi\bi)  at (\bx mm, \by mm) {$X$};
        \node[fill=\colorinter!20!white,  draw=gray,minimum width=\itemw mm, minimum height=\itemh mm] (q\rowi\bi) at (\bx mm, \iy mm) {$Q$};
        \node[fill=\colorinter!30!white,  draw=gray,minimum width=\itemw mm, minimum height=\itemh mm] (o\rowi\bi) at (\bx mm, \qy mm) {$O$};
        \node[fill=\colorkv!50!white,draw=gray,minimum width=\itemw mm, minimum height=12mm,align=center] (k\rowi\bi) at (\kx mm, \ky mm) {$K$ \& $V$};
        \node[fill=gray!20!white,draw=gray,minimum width=\itemw mm, minimum height=\itemh mm] (y\rowi\bi)  at (\bx mm, \oy mm) {$Y$};
    }
    \draw[line width=.8pt, color=\bcolor!50!black,->] (x\rowi\bs) -- (x\rowi\bs |- pa) -- (pa -| k\rowi1) -- (k\rowi1);
    \draw[line width=.8pt, color=\bcolor!50!black,->] (pa) -- (pa -| x\rowi1) -- (q\rowi1);
    \draw[line width=.8pt, color=\bcolor!50!black,->] (o\rowi\bs) -- (pb -| o\rowi\bs) -- (pb -| y\rowi1) -- (y\rowi1);
    \draw[line width=.8pt, color=\bcolor!50!black   ] (pb) -- (pb -| y\rowi1);

    \ifthenelse{\rowi=0}{
        \node[color=\bcolor!20!black,anchor=south east, align=right] at ([xshift=5mm]k\rowi1 |- pa) {\bf Poor GPU\\\bf Utilization};
        \node[anchor=south, align=right] at (x\rowi1.north) {Small Batch};
    }{
        \node[color=\bcolor!20!black,anchor=south east, align=right] at ([xshift=4mm]k\rowi1 |- pa) {Good GPU\\Utilization};
        \node[color=red!20!black,anchor=north, align=center] at ([xshift=2mm,yshift=0mm]k\rowi\bs.south) {\bf Out of GPU\\\bf Memory};
        \node[anchor=west, align=right] at (x\rowi1.east) {Largh Batch};
    }
    
    \draw[line width=.8pt, color=blue!50!black,densely dashed,->] (q\rowi\bs) -- (q\rowi\bs |- k\rowi\bs) -- (k\rowi\bs -| o\rowi1) -- (o\rowi1);
    \draw[line width=.8pt, color=blue!50!black,densely dashed   ] (k\rowi\bs) -- (k\rowi\bs -| o\rowi1);
}

\tikzmath{
    \lx = \paramx * 1.5 + \kvx + 4;
    \ly = \paramy - \yh * 2 + 6;
    \lofst = 1;
    \llen = \lofst + 4;
    \lmid = (\lofst + \llen) / 2;
}

\node[] (tbp) at (\lx mm, \ly mm) {\stype};
\node[] (tsp) at ([yshift=-6mm]tbp) {\btype};

\draw[line width=.8pt, color=red!50!black] ([xshift=-\llen mm]tsp.west) to ([xshift=-\lmid mm]tsp.west);
\draw[line width=.8pt, color=green!50!black  ] ([xshift=-\lmid mm]tsp.west) to ([xshift=-\lofst mm]tsp.west);
\draw[line width=.8pt, color=blue!50!black, densely dashed] ([xshift=-\llen mm]tbp.west) to ([xshift=-\lofst mm]tbp.west);
\draw[line width=.4pt, color=gray] ([xshift=-6 mm, yshift=3 mm ]tbp.west) rectangle ([xshift=1mm, yshift=-3mm]tsp.east);

\end{tikzpicture}
    \caption{Performance dilemma in auto-regressive generation}
    \label{fig:bsissue}
\end{figure}
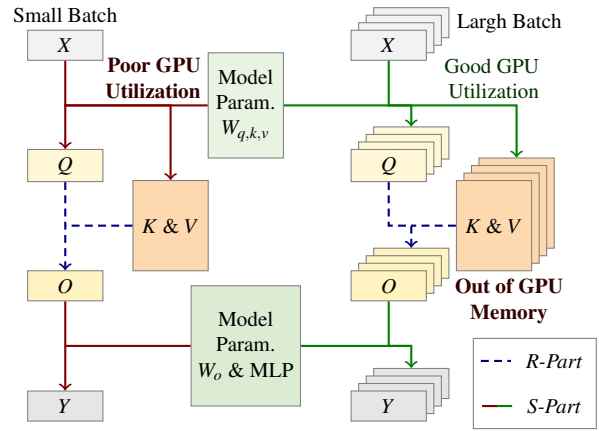

Throughput of the fully connection layers in the transformer blocks increases significantly with a larger batch size.
However, the attention operation benefits little when enlarging the batch size, because each sequence has different $K$ and $V$.
When generating sequences in a batch, instead of GeMM, the GeMV becomes batched GeMV, which is still memory-bound.
To make it worse, the size of \kvcache{} is proportional to the batch size.
As shown in \Cref{fig:bsissue}, when using a larger batch size to better saturate the computation power of GPU, the memory footprint of \kvcache{} is much larger than the capacity of GPU memory.
In fact, the key difficulty is handling the memory-intensive operations with \kvcache{} when using only GPUs for computation.

We categorize the computation workload during generating a token into two parts.
They are denoted using dashed and solid lines, respectively, in \Cref{fig:bsissue}.

\begin{itemize}
    \item \rpart{}: The auto-\textbf{R}egressive computation related to preceding tokens in the sequence, as formulated in \cref{equ:attn} and \cref{equ:weisum}.
Each sequence is processed independently with its own \kvcache{}.
It benefits little from enlarging batch size, but introduces huge memory footprint.
Notably, no model parameter is involved in \rpart{}.
    \item \spart{}: Rest of the model where sequences \textbf{S}hare the same parameters. 
    It mainly consists of fully connected layers. 
    GPU utilization can be significantly increased by batching tokens in more sequences together in \spart{}.
\end{itemize}

\subsection{CPUs can Undertake More in LLM}

In the LLM inference workload, the \kvcache{} takes enormous memory, ideal to be placed in the larger CPU-side DRAM,
despite the bottleneck to move the data from host memory to GPU memory for computation.

However, the \rpart{} is inherently memory-bound workload, where using a GPU gets little benefit over using CPUs.
Therefore, we get our \textbf{key insight}: not only should we store \kvcache{} in CPU memory, but also process them with CPUs.
In other words, \rpart{} should be \textbf{processed near data}.

As a result, the \kvcache{} is removed from GPU memory, and the batch size can be as large as $1024$ or more.
Recall the throughput curve in \Cref{fig:perfmot}. 
The computation in \spart{} is able to utilize the GPUs with much higher efficiency.
The overall token generation throughput is significantly increased.

Two concerns are intuitively raised on such design.
(1) CPUs may not be fast enough to match the throughput of GPUs. 
(2) Transmitting the intermediate data between \spart{} and \rpart{} across devices may be slow.

\begin{table}[h]
    \caption{Latency of Computation Operations}
    \tzfontsize{}
    \centering

\begin{tabular}{ccllll}
\hline
\multirow{2}{*}{Operation}                                                                                 & \multirow{2}{*}{Batch Size} & \multicolumn{2}{c}{Lat. / ms}                     & \multicolumn{2}{c}{TFLOPs}                        \\
                                                                                                           &                             & \multicolumn{1}{c}{GPU} & \multicolumn{1}{c}{CPU} & \multicolumn{1}{c}{GPU} & \multicolumn{1}{c}{CPU} \\ \hline
\multirow{2}{*}{\begin{tabular}[c]{@{}c@{}}\stype{}\\ (\cref{equ:attn} \& \cref{equ:weisum})\end{tabular}} & 1                           & 0.084                   & 0.287                   & 0.050                   & 0.015                   \\
                                                                                                           & 1024                        & 8.32                    & {\ul \textbf{8.12}}     & 0.516                   & 0.529                   \\
\multirow{2}{*}{\begin{tabular}[c]{@{}c@{}}\btype{}\\ ($\sim 16\times$ \cref{equ:out})\end{tabular}}       & 1                           & 1.46                    & 49.5                    & 0.366                   & 0.011                   \\
                                                                                                           & 1024                        & {\ul \textbf{7.08}}     & 611                     & 77.5                    & 0.899                   \\ \hline
\end{tabular}
    \label{table:latcomp}
\end{table}

\Cref{table:latcomp} compares latency of computation tasks in generating tokens on two CPU nodes with those on an A10 GPU.
We highlight the latency of our selected mapping between device and model part using underlines.
When generating tokens using a widely-used 7B foundation model,
the latency of \rpart{} is almost identical between using the GPU or CPUs, as the total hardware memory bandwidths are similar.

When computing on GPUs, the latency of \spart{} is at a similar magnitude with \rpart{}.
However, the \spart{} cannot be moved to CPUs, because it involves much heavier computation that can be greatly accelerated by the GPU.
The poor throughput of CPUs leads to very high latency.
It is also notable as the batch size is $1024\times$ larger, the latency is only $5\times$ larger.
There is more than $100\times$ potential throughput gain.
Overall, with similar latency and throughput of \rpart{}, efficiency is increased in \spart{} and thus the whole system.

\begin{table}[ht]
    \caption{Size of Data and Communication Latency}
    \tzfontsize{}
    \centering

\begin{tabular}{cclll}
\hline
\multirow{2}{*}{Data}                                                                  & \multirow{2}{*}{Batch Size} & \multicolumn{1}{c}{\multirow{2}{*}{Data Size}} & \multicolumn{2}{c}{Latency / ms}                      \\
                                                                                       &                             & \multicolumn{1}{c}{}                           & \multicolumn{1}{c}{PCIe*} & \multicolumn{1}{c}{RoCE*} \\ \hline
Model Weight                                                                           & N/A                         & 402 MB                                         & 12.6                      & 32.2                      \\
\multirow{2}{*}{KV-Cache}                                                              & 1                           & 4.19 MB                                        & 0.131                     & 0.335                     \\
                                                                                       & 1024                        & 4.29 GB                                        & 134                       & 343                       \\ \hline
\multirow{2}{*}{\begin{tabular}[c]{@{}c@{}}Intermediate\\ Vectors (ours)\end{tabular}} & 1                           & 32.7 KB                                        & <0.01                     & 0.03                      \\
                                                                                       & 1024                        & 33.5 MB                                        & 1.04                      & 2.68                      \\ \hline
\end{tabular}

\flushleft
{\footnotesize * Calculated using 32 GB/s PCIe 4.0 x16 and 100 Gbps RoCE}
    \label{table:latcomm}
\end{table}

Sizes of data within a transformer block of a typical 7B model and the latency to send them across different types inter-device connection are shown in \Cref{table:latcomm}.
Instead of previous works that may send the huge model or \kvcache{} across the links, 
we only communicate intermediate vectors, i.e., $Q_i, K_i, V_i, O_i$ in \cref{equ:attn} and \cref{equ:weisum}.
These vectors are orders of magnitudes smaller than others.
The estimated latency to send a large batch of them from $1024$ sequences over the network is only a few milliseconds.
The latency is a moderate portion of the computation latency presented in \Cref{table:latcomp}.
Compared with existing systems, communication overhead of our near-KV-cache processing design is much smaller.

In brief, our approach enables serving LLMs with a much larger batch size.
It brings huge potential of throughput gain, despite the minor overhead that is minimized in our system.
\section{Methodology}
\label{sec:overview}
\subsection{System Overview}

The local CPU in a server equipped with GPU may be too busy and too slow to provide sufficient throughput of processing the \stype{}.
Our approach uses out-of-chassis CPUs, whose aggregated compute power is exploited.

\newcommand{\gpuworker}{S-worker}
\newcommand{\gpuworkers}{S-workers}
\newcommand{\cpuworker}{R-worker}
\newcommand{\cpuworkers}{R-workers}

\begin{figure}[h]
    \centering
    \newcommand{\locallw}[0]{.4}
\newcommand{\commlw}[0]{.8}
\begin{tikzpicture}
\tzfontsize{}

\tikzmath{
    \gsx = 10;
    \gsw = 60;
    \gsy = 40;
    \gsh = 32;
    \gwx = \gsx + 2;
    \gww = \gsw - 4;
    \gwy = \gsy - 4;
    \gwh = \gsh - 5;
    \stw = 10;
    \osx = \gwx + \gww / 2 + 1;
    \osy = \gsy - 1;
}

\draw[color=gray,fill=\colorgpu!50!white,line width=.4] (\gsx mm, \gsy mm) rectangle ++(\gsw mm, -\gsh mm);
\node[anchor=north west] at (\gsx mm, \gsy mm) {GPU Node};
\draw[color=gray,fill=\colorsys!30!white,line width=.4] (\gwx mm, \gwy mm) rectangle ++(\gww mm, -\gwh mm);
\node[anchor=north west] at (\gwx mm, \gwy mm) {\gpuworker{}};

\node[draw=gray,dashed,align=center,anchor=north,fill=white,opacity=.7] (osp) at (\osx mm, \osy mm) {Original\\\stype{} w/\\KV-Cache};

\foreach \gmi in {0, 1}{
    \tikzmath{
        \stm = 9;
        \wx = \gwx + \stm + \gmi * (\gww - 2 * \stm);
        \imw = 6;
        \ix = \wx + (1 - \gmi * 2) * \stw;
        \isx = \ix + (1 - \gmi * 2) * 6;
        \isy = \gwy - \gwh - 3 - \gmi * 3;
        \wy = \gwy - 18;
        \mpy = \wy + 5;
        if \gmi == 0 then {
            \mph = 6;
        } else {
            \mph = 10;
        };
        \mpt = 50 + 20 * \gmi;
    }
    \node[color=white,fill=green!30!black,minimum width=\stw mm] (bp\gmi) at (\wx mm, \wy mm) {\bf \btype{}};
    \node[draw=gray,line width=.4,fill=\colormodel!\mpt!white,anchor=south,align=center,minimum height=\mph mm] (mp\gmi) at (\wx mm, \mpy mm) {Model\\Params.};
    \draw[color=black,line width=.4,->] (mp\gmi) to (bp\gmi);
    
    \node[coordinate] (sp\gmi) at (\isx mm, \isy mm) {};

    \foreach \bi in {0,1,2} {
        \tikzmath{
            \iix = \ix - 1 + \bi;
            \iiy = \wy - 1 + \bi;
            \iit = 10 + \bi * 10;
        }
        \ifthenelse{\gmi=0}{
            \node[draw=gray,fill=\colorinter!\iit!white,minimum width=2 mm,align=center] (qkv\bi) at (\iix mm, \iiy mm) {$Q_i$ \\ $K_i$ \\ $V_i$};
        }{
            \node[draw=gray,fill=\colorinter!\iit!white,minimum width=2 mm] (o\bi) at (\iix mm, \iiy mm) {$O_i$};
        }
    }
    \ifthenelse{\gmi=0}{
        \draw[color=black,line width=\locallw,->] ([xshift=-3mm]bp\gmi.west) to (bp\gmi);
        \draw[color=black,line width=\locallw,->] (bp\gmi) to (qkv0.west |- qkv1);
        \draw[color=black,line width=\commlw,color=purple!50!black] (qkv1 -| qkv2.east) -- (qkv1 -| sp\gmi) -- (sp\gmi);
        \draw[color=gray,dashed,->] (qkv2) -- (qkv2 |- osp) -- (osp);
    }{
        \draw[color=black,line width=\locallw,->] (bp\gmi) to ([xshift=3mm]bp\gmi.east);
        \draw[color=black,line width=\locallw,->] (o2.east |- o1) to (bp\gmi);
        \draw[color=black,line width=\commlw, ->,color=brown!50!black] (sp\gmi) -- (o1 -| sp\gmi) -- (o0.west |- o1);
        \draw[color=gray,dashed,->] (osp) -- (o2 |- osp) -- (o2);
    }
}

\tikzmath{
    \ch = 32;
    \cy = \gsy - \gsh - 8 - \ch;
    \cxmargin = 4;
    \cw= 24;
    \cxo = 40 - \cw * 1.5 - \cxmargin;
}

\foreach \cwi in {0, 1, 2}{
    \tikzmath{
        \cwn = int(\cwi + 1);
        \cix = \cxo + \cwi * (\cxmargin + \cw);
        \cwx = \cix + 1;
        \cww = \cw - 2;
        \cwy = \cy + 4;
        \cwh =  \ch - 5;
        \cwt = 20 + 10 * \cwi;
    }
    \draw[color=gray,fill=\colorcpu!50!white,line width=.4] (\cix mm, \cy mm) rectangle ++(\cw mm, \ch mm);
    \node[anchor=south west] at (\cix mm, \cy mm) {CPU Node \cwn};
    \draw[color=gray,fill=\colorsys!\cwt!white,line width=.4] (\cwx mm, \cwy mm) rectangle ++(\cww mm, \cwh mm);
    \node[anchor=south west] at (\cwx mm, \cwy mm) {\cpuworker{} \cwn};
    \tikzmath{
        \iit = 10 + \cwi * 20;
        \qx = \cwx + 1;
        \qy = \cwy + \cwh - 6;
        \ox = \cwx + \cww - 1;
        \px = \cwx + \cww - 10;
        \py = \qy - 4;
        \kvx = \px + 2;
        \kvy = \py - 9;
    }
    \node[draw=gray,line width=\commlw,fill=\colorinter!\iit!white,anchor=south west,align=center,minimum width=5mm] (q\cwi) at (\qx mm, \qy mm) {$Q_i$};
    \node[draw=gray,line width=\commlw,fill=\colorinter!\iit!white,anchor=north west,align=center,minimum width=5mm] (kv\cwi) at (\qx mm, \qy mm) {$K_i$\\$V_i$};
    \node[draw=gray,line width=\commlw,fill=\colorinter!\iit!white,anchor=south east,align=center,minimum width=5mm] (oc\cwi) at (\ox mm, \qy mm) {$O_i$};
    \ifthenelse{\cwi=1}{
        \draw[line width=\commlw,color=purple!50!black,->] (sp0 -| q\cwi) to (q\cwi.north);
        \draw[line width=\commlw,color= brown!50!black] (oc\cwi) -- (oc\cwi |- sp1);
    }{
        \draw[line width=\commlw,color=purple!50!black,->] (sp0) -- (sp0 -| q\cwi) to (q\cwi.north);
        \draw[line width=\commlw,color=brown!50!black] (oc\cwi) -- (oc\cwi |- sp1) -- (sp1);
    }
    \node[circle,fill=purple!50!black,minimum width=1mm,minimum height=1mm,scale=.5] at (sp0 -| q\cwi) {};
    \node[circle,fill=brown!50!black,minimum width=1mm,minimum height=1mm,scale=.5] at (sp1 -| oc\cwi) {};
    
    \node[color=white,fill=blue!20!black,minimum width=\stw mm] (stype\cwi) at (\px mm, \py mm) {\bf \stype{}};
    \node[draw=gray,line width=.4,fill=\colorkv!\iit!white,align=center] (kvc\cwi) at (\kvx mm, \kvy mm) {KV-Cache\\Part. \cwn};
    \draw[color=black,line width=.4,->] (q\cwi) -- (stype\cwi |- q\cwi) -- (stype\cwi);
    \draw[color=black,line width=.4,->] (stype\cwi) -- (stype\cwi -| oc\cwi) -- (oc\cwi);
    \draw[color=black,line width=.4,->] (kv\cwi) -- (kv\cwi |- kvc\cwi) -- (kvc\cwi);
    \draw[color=black,line width=.4,->] (kvc\cwi.north-|stype\cwi) -- (stype\cwi);
}

\node[diamond,fill=purple!50!black,minimum width=1mm,minimum height=1mm,scale=.5] at (sp0) {};
\node[diamond,fill=brown!50!black,minimum width=1mm,minimum height=1mm,scale=.5] at (sp1) {};

\end{tikzpicture}
    \caption{Workers of \sys{}}
    \label{fig:overview}
\end{figure}

\Cref{fig:overview} shows the basic design of \sys{}, which consists of two types of workers.

An \textit{\gpuworker{}} computes \btype{} of an LLM.
It may use one or multiple GPUs.
All weights of the model are on the \gpuworker{}, and partitioned by a certain way of model parallelism if using multiple GPUs.
It acts as a typical token generation worker simply using GPUs, except for its much larger batch size and the behavior of computing \stype{}.
To generate a new token, it goes through the transformer blocks.
After $Q_i, K_i, V_i$ are produced by fully connected layers in \btype{},
instead of computing \stype{} locally,
the \gpuworker{} sends different parts of them related to different sequences to the \cpuworkers{},
and retrieve the output, $O_i$, from them.
Then, it feeds $O_i$ to succeeding layers in \btype{} on the GPU.

The \textit{\cpuworkers{}} use CPUs on remote nodes to compute \stype{}.
These \cpuworkers{} are light-weight, because no model parameter is involved in \stype{} of LLMs.
The functionality of a \cpuworker{} is simple.
It receives $Q_i, K_i, V_i$ of a batch of tokens.
$K_i$ and $V_i$ are appended to the existing KV-cache.
$Q_i$ is used to in attention computation with the local KV-cache data, and the output is returned.
The \cpuworkers{} may also drop KV-cache of a certain sequence upon its generation ends.

As GPU is the most scarce resource, the system maximizes the throughput of the \gpuworker{} via maximizing the batch size.
As \kvcache{} is excluded from the \gpuworker{}, there is little tension on GPU memory capacity.
Beside the model weight, it only needs memory for a small scratchpad of the current layer.
The possible batch size can be up to millions of sequences. 
While a larger batch leads to larger latency, we enable the possibility to choose one that best fits the use case.

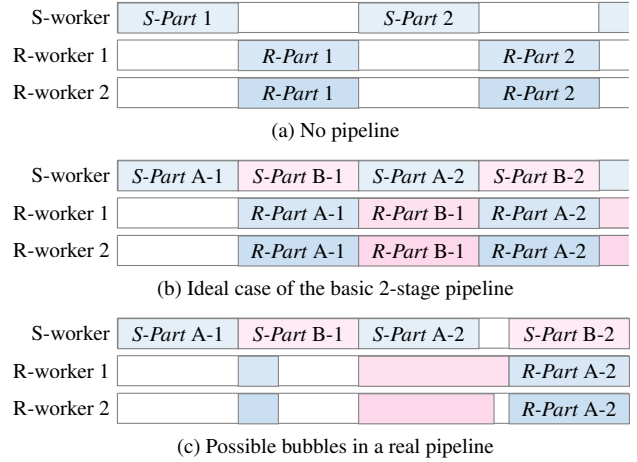
\begin{figure}[h]
    \centering
    \begin{tikzpicture}
\tzfontsize{}

\tikzmath{
    \tlx = 20;
    \tlw = 68;
    \tlh = 4;
    \tlmargin = 1;
    \tlgh = \tlh * 3 + \tlmargin * 2 + 7;
    \rtw  = 16;
}

\def\labels{{
"(a) No pipeline",
"(b) Ideal case of the basic 2-stage pipeline",
"(c) Possible bubbles in a real pipeline"
}}

\foreach \tgi in {0, 1, 2} {
    \tikzmath{
        \by = 50 - \tgi * \tlgh;
        \labely = \by - 3 * \tlh - 3 * \tlmargin;
        \labelx = \tlx - 5 + \tlw / 2;
    }
    \node[anchor=north] at (\labelx mm, \labely mm) {\pgfmathparse{\labels[\tgi]} \pgfmathresult};
    \foreach \ti in {0, 1, 2} {
        \tikzmath{
            \ty = \by - \ti * (\tlh + \tlmargin);
            \tit = \ti * 20 + 40;
            \fnx = \tlx + \rtw * 4;
            \fnw = \tlw - \rtw * 4;
        }
        \draw[color=gray,line width=.4] (\tlx mm, \ty mm) -- ++(\tlw mm, 0);
        \draw[color=gray,line width=.4] (\tlx mm, \ty mm) -- ++(0, -\tlh mm) -- ++(\tlw mm, 0);
        \ifthenelse{\ti=0}{
            \node[anchor=north east] at (\tlx mm, \ty mm) {\gpuworker{}};
        }{
            \node[anchor=north east] at (\tlx mm, \ty mm) {\cpuworker{} \ti};
        }
        \ifthenelse{\tgi < 2}{
            \ifthenelse{\ti=0}{
                \fill[color=\colorba!\tit!white] (\fnx mm, \ty mm) rectangle ++(\fnw mm, -\tlh mm);
                \draw[color=gray,line width=.4] (\fnx mm, \ty mm) -- ++(0, -\tlh mm) -- ++(\fnw mm, 0) (\fnx mm, \ty mm) -- ++(\fnw mm, 0);
            }{
                \ifthenelse{\tgi>0}{
                    \fill[color=\colorbb!\tit!white] (\fnx mm, \ty mm) rectangle ++(\fnw mm, -\tlh mm);
                    \draw[color=gray,line width=.4] (\fnx mm, \ty mm) -- ++(0, -\tlh mm) -- ++(\fnw mm, 0) (\fnx mm, \ty mm) -- ++(\fnw mm, 0);
                }{}
            }
        }{}
        \foreach \li in {0, 1}{
            \tikzmath{
                \lidx = int(\li + 1);
                if \tgi == 0 then { let \tgtxt = \lidx;  } else {
                    let \tgtxt = A-\lidx;
                    let \thtxt = B-\lidx;
                };
                \stg = \tlw - \rtw * 4;
                if \ti == 0 then {
                    \lx = \tlx + \li * 2 * \rtw;
                } else {
                    \lx = \tlx + (\li * 2 + 1) * \rtw;
                };
                \nx = \lx + \rtw;
                if \tgi * 10 + \li == 21 then {
                    if \ti == 0 then { \nx = \nx + \stg; } else { \lx = \lx + \stg; };
                };
                \shb = 0;
                if \tgi * 100 + \li == 200 then { \shs = 1; } else { \shs = 0; };
                \stw = \rtw;
                \htw = \stw / 3;
                \ltw = \stw + \stg * (3 - \ti) / 2;
            }
            \ifthenelse{\ti=0}{
                \ifthenelse{\shb=1}{
                    \node[minimum height=\tlh mm, minimum width=\htw mm,draw=gray,line width=.4,fill=\colorba!\tit!white,anchor=north west] at (\lx mm, \ty mm) {};
                }{
                    \node[minimum height=\tlh mm, minimum width=\stw mm,draw=gray,line width=.4,fill=\colorba!\tit!white,anchor=north west] at (\lx mm, \ty mm) {\spart{} \tgtxt};
                }
            }{
                \ifthenelse{\shs=1}{
                    \node[minimum height=\tlh mm, minimum width=\htw mm,draw=gray,line width=.4,fill=\colorba!\tit!white,anchor=north west] at (\lx mm, \ty mm) {};
                }{
                    \node[minimum height=\tlh mm, minimum width=\stw mm,draw=gray,line width=.4,fill=\colorba!\tit!white,anchor=north west] at (\lx mm, \ty mm) {\rpart{} \tgtxt};
                }
            }
            \ifthenelse{\tgi=0}{}{
                \ifthenelse{\ti=0}{
                    \ifthenelse{\shb=1}{
                        \node[minimum height=\tlh mm, minimum width=\htw mm,draw=gray,line width=.4,fill=\colorbb!\tit!white,anchor=north west] at (\nx mm, \ty mm) {};
                    }{
                        \node[minimum height=\tlh mm, minimum width=\stw mm,draw=gray,line width=.4,fill=\colorbb!\tit!white,anchor=north west,inner sep=0] at (\nx mm, \ty mm) {\spart{} \thtxt};
                    }
                }{
                    \ifthenelse{\li=0}{
                        \ifthenelse{\shs=1}{
                            \node[minimum height=\tlh mm, minimum width=\ltw mm,draw=gray,line width=.4,fill=\colorbb!\tit!white,anchor=north west] at (\nx mm, \ty mm) {};
                        }{
                            \node[minimum height=\tlh mm, minimum width=\stw mm,draw=gray,line width=.4,fill=\colorbb!\tit!white,anchor=north west,inner sep=0] at (\nx mm, \ty mm) {\rpart{} \thtxt};
                        }
                    }{}
                }
            }
        }
    }
}

\end{tikzpicture}

\vspace{-1em}
    \caption{Temporal view of \sys{}}
    \label{fig:basicpipeline}
\end{figure}

In this system, the \gpuworker{} and the \cpuworkers{} work in turns to generate a token.
When one type of worker is working, the other idles, as shown in \Cref{fig:basicpipeline}(a).

Therefore, a basic two-stage pipeline at token level is applied.
As \Cref{fig:basicpipeline}(b) shows, the \gpuworker{} starts with two separate mini-batches: A and B.
After it finishes the first \btype{} of mini-batch A, it starts working on the first \btype{} of mini-batch B.
At the same time, the \cpuworkers{} are working on the \stype{} of mini-batch A.
Overall, the two mini-batches are processed by each type of worker in turns.

This token-level pipeline only achieves optimal utilization of all workers if computation latency of \btype{} and \stype{} are equal.
Otherwise, there are still bubbles in the pipeline due to idling workers, as \Cref{fig:basicpipeline}(c) suggests.
In fact, the pipeline can barely be free of bubbles, because the workload and the hardware are both of high heterogeneity.
The following two techniques addresses the challenges at temporal and inter-device scopes, respectively.

\subsection{Sequence-level Load-Stabilizing Schedule}

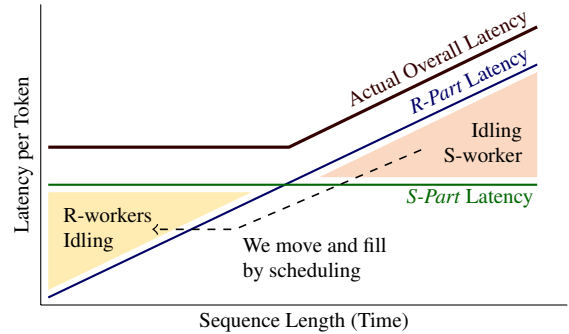
\begin{figure}[ht]
    \centering
    \begin{tikzpicture}
\tzfontsize{}

\tikzmath{
    \ex = 70;
    \ey = 40;
    \hx = \ex / 2;
    \hy = \ey / 2;
    \sx = 1;
    \sy = 1;
    \ux = \ex - 4;
    \uy = \ey - 8;
    \hux = \ux / 2;
    \huy = \uy / 2;
    \lh = 5;
    \hly = \huy + \lh;
    \ly = \uy + \lh;
    \lxs = 1;
    \lys = 2;
}

\draw[line width=.4,color=black] (0, 0) -- (\ex mm, 0) (0, 0) -- (0, \ey mm);
\node[anchor=north] at (\hx mm, 0) {Sequence Length (Time)};
\node[anchor=south, rotate=90] at (0, \hy mm) {Latency per Token};

\draw[line width=0.8, color=blue!40!black] (1mm, 1mm)      -- node[sloped, pos=1., anchor=south east, inner sep=1] {\rpart{} Latency} (\ux mm, \uy mm);
\draw[line width=0.8, color=green!40!black] (1mm, \huy mm) -- node[sloped, pos=1,  anchor=north east, inner sep=1] {\spart{} Latency} (\ux mm, \huy mm);
\draw[line width=1.2, color=red!20!black] (1mm, \hly mm) --  (\hux mm, \hly mm) -- node[sloped, above,inner sep=1, anchor=south east, pos=1] {Actual Overall Latency} (\ux mm, \ly mm);

\fill[color=colorB!50!white] (1mm, 2mm) -- ([yshift=-1mm]1mm, \huy mm) -- ([xshift=-5mm, yshift=-1mm]\hux mm, \huy mm);
\fill[color=colorA!50!white] ([yshift=1mm]\ux mm, \huy mm) -- ([yshift=-1mm]\ux mm, \uy mm) -- ([xshift=4mm, yshift=1mm]\hux mm, \huy mm);

\node[anchor=north west,align=left] (irw) at ([xshift=\lxs mm, yshift=-\lys mm]1mm, \huy mm) {\cpuworkers{}\\Idling};
\node[anchor=south east,align=right] (isw) at ([xshift=-\lxs mm, yshift=\lys mm]\ux mm, \huy mm) {Idling\\\gpuworker{}};

\draw[line width=.6, dashed, ->] ([xshift=-2mm, yshift=-1mm]isw.west) -- ([xshift=10mm]irw.east) node[anchor=north west,align=left] {We move and fill\\by scheduling} -- ([xshift=-1mm]irw.east);

\end{tikzpicture}
    \caption{Hardware idling due to workload heterogeneity}
    \label{fig:slidle}
\end{figure}

There is a significant amount of bubbles in the basic two-stage pipeline,
because the workloads of \rpart{} and \spart{} change differently depending on the length of the generated sequence.
Latency of \spart{} is only related to the batch size, because each new token involves fixed amount of computation in the fully-connected layers.
In contrast, latency of \rpart{} is related to the total length of the sequences, because each new token interacts with all previous tokens in the sequence.
When processing a batch of sequences, the generated sequences get longer over time.
Because the overall latency of \sys{} is dominated by the larger latency of \spart{} and \rpart{}, either \gpuworker{} or \cpuworker{} idles a lot, as shown in \Cref{fig:slidle}.

While changing the workload of \spart{} may lead to decreased GPU utilization,
the overall performance can be improved by reducing the latency fluctuation of \rpart{} without much efficiency issue.
Ideally, as indicated in \Cref{fig:slidle}, we may move the workload of \rpart{}, keeping its total amount unchanged.
The triangular area of idling \gpuworker{} is moved to the area of idling \cpuworkers{}.
As the total overall latency can be indicated by the area under the latency curve, this can reduce as much as $20\%$ the total overall latency.
In other words, the overall throughput is increased by $20\%$.
Besides, the maximum latency to generate a token is reduced by $50\%$, indicated by the highest point of the latency curve.

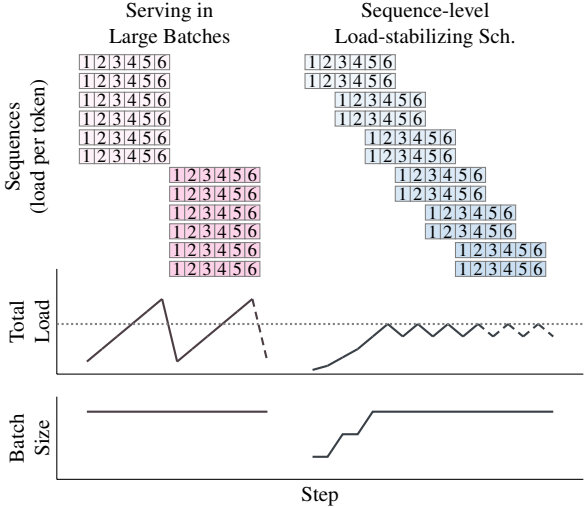
\begin{figure}[ht]
    \centering
    \begin{tikzpicture}
\tzfontsize{}

\tikzmath{
    \ex = 70;
    \ey = 14;
    \hx = \ex / 2;
    \hy = \ey / 2;
    \iw = 2;
    \imargin = \iw + 0.5;
    \sl = 6;
    \bsy = -5;
}

\draw[line width=.4,color=black]
(0, -\ey mm) -- node[midway, below, sloped]{Step} (\ex mm, - \ey mm)
(0, 0) -- (\ex mm, 0)
(0, 0) -- node[midway, above, sloped, align=center]{Total\\Load} (0, \ey mm)
(0, - \ey mm) -- node[midway, above, sloped, align=center]{Batch\\Size} (0, -3 mm);

\def\labels{{
"Serving in\\Large Batches",
"Sequence-level\\Load-stabilizing Sch."
}}

\def\ys{{
    {3, 6, 9, 12, 15, 18, 3, 6,  9, 12, 15, 18},
    {1, 2, 4, 6,  9,  12, 9, 12, 9, 12,  9, 12}
}}

\foreach \chi in {0, 1}{
    \tikzmath{
        \uy = 5 * \imargin + \ey;
        if \chi == 0 then {
            \mx = 5.5;
            let \col = \colorbb;
            \sl = 11;
        } else {
            \mx = 7.5;
            let \col = \colorba;
            \sl = 15;
        };
        \cx = 4 + 30 * \chi;
        \ux = \cx + \iw * \mx;
        \uy = 11 * \imargin + \ey + 1;
    }
    \node[anchor=south,align=center] at (\ux mm, \uy mm) {\pgfmathparse{\labels[\chi]} \pgfmathresult};
    \foreach \si in {0,...,11}{
        \tikzmath{
            if \chi == 0 then {
                \bo = int(\si / 6) * 6;
                \tpt = \bo * 10 + \si * 1 + 20;
            } else { 
                \bo = int(\si / 2) * 2;
                \tpt = \bo * 5 + \si + 20;
            };
            \bx = \cx + \bo * \iw;
            \by = \uy - 1 - \si * \imargin;
        }
        \foreach \ti in {0,...,5}{
            \tikzmath{
                \tx = \bx + \ti * \iw;
                \tidx = int(\ti + 1);
            }
            \node[line width=.4, draw=gray, fill=\col!\tpt!white, inner sep=0, minimum width=\iw mm, minimum height=\iw mm] at (\tx mm, \by mm) {\scriptsize \tidx};
        }
    }
    \foreach \xi in {0,...,\sl}{
        \tikzmath{
            \px = \cx + \xi * \iw;
            \qx = \px + \iw;
        }
        \ifthenelse{\xi<11}{
            \tikzmath{
                \py = (\ey - 4) * \ys[\chi][\xi] / 18;
                \qy = (\ey - 4) * \ys[\chi][\xi + 1] / 18;
            }
            \draw[line width=.8, color=\col!30!black] (\px mm, \py mm) -- (\qx mm, \qy mm);
        }{
            \tikzmath{
                \py = (\ey - 4) * \ys[\chi][\xi - 6] / 18;
                \qy = (\ey - 4) * \ys[\chi][\xi - 5] / 18;
            }
            \draw[line width=.8, color=\col!30!black, densely dashed] (\px mm, \py mm) -- (\qx mm, \qy mm);
        }
    }
    \ifthenelse{\chi=0}{
        \tikzmath{
            \zx = \cx + (\sl + 1) * \iw;
        }
        \draw[line width=.8, color=\col!30!black] (\cx mm, \bsy mm) -- (\zx mm, \bsy mm);
    }{
        \tikzmath{
            \xy = \bsy - (\bsy + \ey) / 3 * 2;
            \yy = \bsy - (\bsy + \ey) / 3 * 1;
            \xx = \cx + 2 * \iw;
            \yx = \cx + 4 * \iw;
            \zx = \cx + (\sl + 1) * \iw;
        }
        \draw[line width=.8, color=\col!30!black] (\cx mm, \xy mm) -- ([xshift=\iw mm]\cx mm, \xy mm) -- (\xx mm, \yy mm) -- ([xshift=\iw mm]\xx mm, \yy mm) -- (\yx mm, \bsy mm) -- (\zx mm, \bsy mm);
    }
}

\tikzmath{
    \zy = 12 / 18 * (\ey - 4);
    \ly = \ey + 6.5 * \imargin;
}
\draw[densely dotted, color=gray, line width=.8] (0, \zy mm) -- (\ex mm, \zy mm);
\node[rotate=90, anchor=south, align=center] at (0, \ly mm) {Sequences\\(load per token)};

\end{tikzpicture}
    \caption{The sequence-level load-stabilizing schedule}
    \label{fig:slpipe}
\end{figure}

We identify that the long latency of \rpart{} is caused by all sequences being long.
As previous works~\cite{orca} indicates, sequences of different lengths can be batched together in \spart{} to increase throughput.
And in \rpart{}, processing the sequences separately on different workers introduce no extra overhead.
Therefore, we schedule the sequences in a pipeline to control the total length being processed at each step, as shown in \Cref{fig:slpipe}.

\newcommand{\launchinterval}{F}

Instead of starting a large batch of sequences together, smaller micro-batches are started with a fixed interval, $\launchinterval$ steps.
As a sequence of target length $S$ is generated in $S$ steps, multiple micro-batches are being processed together in each step.
The total size of the micro-batches being concurrently processed roughly equals to the total batch size.
\spart{} of these micro-batches are processed together as a large batch, so the GPUs are well-utilized in the same way as serving the sequences in a large batch. 

For example, in \Cref{fig:slpipe}, size of the micro-batch is $2$. 
After 4 steps of cold starting, the total number of sequences to be processed at each step equals to $6$, the original large batch size.
The sum of the numbers in a column indicates the total load in \rpart{} of the step.
In the original large-batch schedule, it can be as much as $36$ in a step, while the number is $24$ in the load-stabilizing schedule, indicating 1/3 reduction of the maximum latency.
In fact, the reduction of latency can be nearly halved, following the formal deduction below.

This schedule reduces the total length of all sequences by mixing sequences of different lengths together.
To be more specific, assume that there are originally $B$ sequences of length $S$.
The total length of sequences can be $W_{\max} = BS$ in the final step if all sequences are started together.
In our schedule, the size of micro-batches is defined as follows.

\begin{equation}
    M = \frac{B\launchinterval}{S}
\end{equation}

So, in the final step of generating a micro-batch, we have the maximum total sequence length.

\begin{equation}
    W'_{\max} = \sum_{k=1}^{S/\launchinterval} M k \launchinterval  = \frac{B(S+\launchinterval)}{2} \approx \frac{BS}{2} = \frac{W_{\max}}{2}
\end{equation}

Although $S=\frac{1}{3}\launchinterval$ in the example in \Cref{fig:slpipe} leads to $W'_{\max} = \frac{2}{3} W_{\max}$, $S$ is usually much larger than $\launchinterval$ in real cases (thousands compared to tens). 
So, $(S+\launchinterval)$ is closer to $S$.
As the approximation indicates, the maximum total length of sequences is reduced by $50\%$.

The total workload remains near $W'_{\max}$ throughout the rest contiguous serving process. 
Because there are almost infinite number of micro-batches to be processed, 
after the cold start of the pipeline schedule, there are always sequences of different lengths being processed.

An extra benefit is the reduction of waiting time for incoming sequence generating requests in online serving.
It has to wait for up to $S$ steps before being served in a large batch, but only waits for $\launchinterval$ steps in the load-stabilizing schedule.

\begin{algorithm}[h]
    \caption{Load-control Algorithm}
    \label{algo:loadcontrol}
        \begin{algorithmic}[1]
        \Require{$M$: array of batch sizes of all current micro-batches}
        \Require{$E$: ending step index of all current micro-batches}
        \Require{$W$: array of workload}
        \Require{$t$: starting step index of the micro-batch}
        \Require{$m$: size of the micro-batch}
        \Function{AddMicroBatch}{}
            \State $M$.append$(m)$
            \State $E$.append$(t + S)$
            \State $W$.append$(m * S)$
            \ForAll{$i$ \textbf{in} current micro-batches}
                \State $W[i] \gets W[i] + (E[i] - t) * m$
            \EndFor
        \EndFunction
        
        \Function{GetEariliestStep}{}
            \State $r \gets t$
            \ForAll{$i$ \textbf{in} current micro-batches}
                \State $x \gets \lfloor \frac{W_{\lim} - W[i]}{m} \rfloor$ \Comment{Maximum allowed length}
                \State $r \gets \max (r, E[i] - x + 1)$
            \EndFor
            \State \Return $r$
        \EndFunction
    \end{algorithmic}
\end{algorithm}

The sequence-level load-stabilizing schedule can be generalized to a load control algorithm that dynamically determines when a new micro-batch starts.
Given a maximum load limit $W_{\lim}$, the earliest starting step index for a micro-batch of $M$ sequences can be calculated based on the current micro-batches being processed.
\Cref{algo:loadcontrol} shows the algorithm that figures out the earliest step index, with a few more information to maintain when a micro-batch actually starts.
As maximum total length is reached at the last step of each micro-batches, the algorithm maintains the workload at these steps. 
The margin between the maximum workload and the limit is used to get the maximum length of the new micro-batch at the specific step,
Then, we get the earliest step index constrained by a certain peak of workload.

Notably, during cold starting of the schedule, there may be an issue to set $W_{\lim}$ to $W'_{\max}$.
With sufficient input to be processed, a large number of sequences are started at step $0$.
Then, step $S$ becomes the peak, and the rest sequences can only launch at step $S+1$.
Instead, we need to gradually increase $W_{\lim}$, or use a fixed $\launchinterval$ in the beginning.
\subsection{Workload-balanced Hardware Selection}

\sys{} introduces hardware heterogeneity between the GPU and CPUs.
To optimally utilize both of them, beside stabilizing the workload, selection of hardware also makes significant impact.
Specifically, we need to determine the number of CPUs to use in our system.
If we have too many CPUs, it is a waste because they have to wait for the GPU.
If we have too few CPUs, it is the GPU that idles.

Also, the expected service latency to LLM users should be considered.
In some cases, the acceptable latency to generate a single sequence is large, so we have more space to optimize the throughput.
In other cases, the batch size should be reduced to fulfill a stricter latency limit.

\newcommand{\batchsize}{\mathcal{B}}
\newcommand{\ncpus}{\mathcal{P}}

We introduce a quantitative approach to determine the two most important parameters of our system: the batch size, $\batchsize$, and the number of CPUs, $\ncpus$.

To start with, there are two given conditions: the LLM and the GPU we use.
Then, we need a few reference metrics.

\newcommand{\latfunc}{\mathbb{T}(\batchsize)}

Throughput to compute \btype{} of the model on the GPU can be measured by a micro-benchmark.
As shown in \Cref{fig:perfmot}, the throughput varies significantly as the batch size, $\batchsize$, changes.
Therefore, we use a function $\latfunc$ to indicate the latency to compute \btype{} of one transformer block on the GPU.
Also, we have the user-specified expected maximum length of sequences $S$.

We assume that perfect efficiency is achieved by the pipelines.
As the latency of \btype{} is fixed, and the latency of \stype{} equals to the latency of \btype{} in such a pipeline,
the latency to generate a token using a model of $N$ layers is calculated as follows.

\begin{equation}
\label{equ:bscon}
    2 N S \cdot \latfunc \leq L
\end{equation}

\newcommand{\latfuncper}{\mathbb{E}(\batchsize)}

$L$ indicates the expected latency to generate a sequence.
Larger $\batchsize$ leads to larger $\latfunc$, as well as better overall throughput.
A maximum possible $\mathcal{B}$ is selected as the above constraint is fulfilled.

If there is no constraint on $L$, $\batchsize$ is selected based on the overall throughput on GPU.

\begin{equation}
    \latfuncper = \frac{\batchsize}{\latfunc}
\end{equation}

$\latfuncper$ is proportional to the GPU throughput that is shown in \Cref{fig:perfmot}.
It increases sharply when $\batchsize$ is small, indicating that increasing $\batchsize$ brings much benefit.
When it becomes more stable as $\batchsize$ is large enough, the performance gains little.
In this case, we should select a $\batchsize$ that further increasing it only brings marginal throughput improvement.

Another constraint of $\batchsize$ is the host-side memory capacity.
Assume that each CPU has memory for $K$ and $V$ vectors for $C$ tokens, which can be calculated from the size of the memory and specifications of the model.

\begin{equation}
\label{equ:memcon}
    \frac{1}{2} \batchsize S \leq C\ncpus
\end{equation}

In fact, this constraint is barely the actual limitation, because CPUs commonly have abundant memory.

After having $\batchsize$ determined, $\ncpus$ is minimized by the constraint of computing \stype{} in similar time to \btype{}.
Assume that we are using the same model of CPUs in the system.
We use another micro-benchmark to get $R$, which indicates the latency that one CPU processes one token for \stype{}.
So, the CPU takes time $Rk$ in \stype{} when generating a token appending to a sequence of $k$ existing tokens.
We get the constraint for $\ncpus$ using $R$.

\begin{equation}
    \frac{\batchsize S}{2\ncpus} R \approx \latfunc
\end{equation}

Then, we get a direct approximation for the optimal number of CPUs to work with a GPU.

\begin{equation}
\label{equ:cpucon}
    \ncpus \approx \frac{\batchsize S R}{2 \latfunc} = \frac{1}{2} SR \latfuncper
\end{equation}

Briefly, to cope with increased GPU efficiency $\latfuncper$ thanks to increased $\batchsize$, more CPUs are needed.
However, as we select the $\batchsize$ where increasing it brings marginal $\latfuncper$, and $\ncpus$ has to be an integer, it has little impact when tweaking $\batchsize$.
Also, longer expected sequence length $S$ makes the CPUs more heavily loaded, so more of them are needed.

In summary, given specifications of the hardware and the model, we first measure $\latfunc$ and $R$ with micro-benchmarks.
Then, a definite optimal choice of $\batchsize$ and $\ncpus$ is given by \Cref{equ:bscon}, \Cref{equ:memcon}, and \Cref{equ:cpucon}.

Furthermore, assume that the feature dimension of the model is $h$.
The workload of \spart{}, reflected in $\latfunc$, is proportional to $h^2$.
Meanwhile, $R$, the per-token workload of \rpart{}, is proportional to $h$.
So, $\ncpus$ is approximately proportional to $\frac{1}{h}$.
The optimal number of CPUs tends to be smaller for larger $h$, which commonly appears in larger models.

\section{Implementation}
\label{sec:impl}

The \gpuworker{} of \sys{} is implemented using PyTorch, for ease of adapting to various models and serving APIs.
\stype{} is stripped from the model and the token generation scheduling is taken over by our system.
The \cpuworker{} is implemented using C++.
As a light-weight service, it receives data from \gpuworker{}.
We find that the performance the \cpuworker{} is more critical but understudied.

\subsection{Mix-precision CPU Attention}

Optimizing the performance of \cpuworker{} is critical to the overall throughput of the system.
Compared to the well-established neural network libraries on GPUs, there lacks existing high performance neural network libraries on CPUs that can be used out-of-box.
Most current LLMs use 16-bit floating point numbers (fp16), which is not supported by most CPU libraries.
However, using fp32 libraries doubles the volume of memory access, which means doubling the latency.

We develop a mixed-precision attention operator that reads fp16 data from memory, convert them to fp32 in registers, and compute.
Luckily, we find intrinsics in AVX-2 instruction set to perform the vectorized fp16-fp32 conversion in one instruction.
Although fp16 floating point multiply and add (FMA) instruction is included in AVX-512 instruction set, we exclude it for compatibility to a wider range of CPUs.

\subsection{Supporting Quantization}

The above fp16-fp32 mix-precision implementation is lossless comparing with the original fp16 computation on GPUs.
We also support more aggressive performance optimization if model accuracy degradation is tolerated.
Model quantization is widely used and welcomed to boost our performance.

Various quantization algorithms are supported with a few extra functions to implement.
Given $Q_i, K_i, V_i$ vectors in fp16, the user function adds $K_i$ and $V_i$ to the \kvcache{} after quantization.
$Q_i$ is transformed as the quantization algorithm requires to produce $O_i$.

Our throughput benefits from storing the \kvcache{} data in a quantized format.
Suppose that 4-bit integers are used to store $K$ and $V$, the memory access size is quartered, and we are likely to get $4\times$ speed up, or save $4\times$ CPUs.

\subsection{Model Parallelism}

For larger LLMs, model parallelism at either tensor or layer level is a common technique that reduces the computation latency.
It is mandatory in many cases as the model cannot fit in the memory of a single GPU.

\sys{} naturally have good support for such techniques.
A separate group of \cpuworkers{} are assigned to every \gpuworker{} that plays the part of a worker in the parallel groups.

For inter-layer model parallelism, i.e., pipeline parallelism, different transformer blocks are processed by each worker.
So, \rpart{} related to each worker are totally independent.

For intra-layer model parallelism, i.e., the tensor-model parallelism, the fully-connected layers before and after \rpart{} are commonly partitioned across attention heads~\cite{megatron}.
Therefore, each group of \cpuworkers{} maintains independent \kvcache{} for different attention heads.

In both types of model parallelism, the workloads of \spart{} and \rpart{} are divided by a same factor.
Therefore, the number of \cpuworkers{} to work with each \gpuworker{} remain unchanged, as revealed in \Cref{equ:cpucon}.
Latency of \sys{} directly benefits from the resulting latency reduction.
\section{Evaluation}
\label{sec:eval}

\subsection{Setup}

\paragraph{Models and tasks}
All the auto-regressive models use the same transformer backbone.
We choose a state-of-the-art open-source LLMs, Llama~\cite{llama2} and OPT~\cite{opt}. 
We evaluate system performance over different sizes of the models, including Llama-7b, Llama-13b, and Opt-175b. 

\begin{figure}[ht]
    \centering
    \includegraphics[scale=.5]{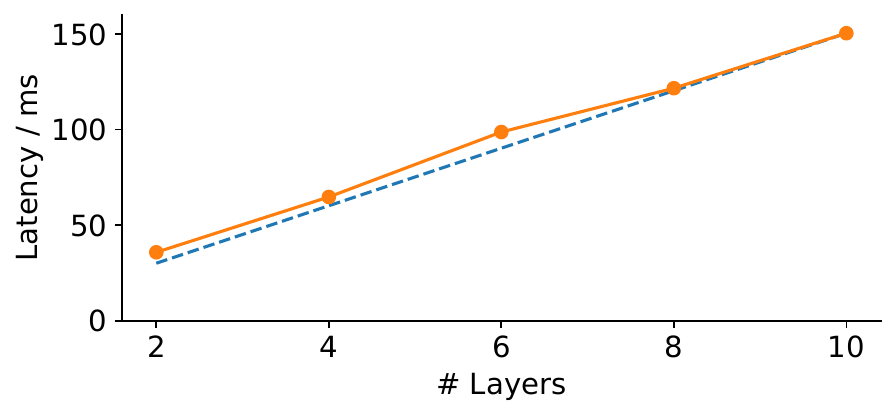}
    \caption{Latency of \sys{} serving Opt-175b}
    \label{fig:layerlat}
\end{figure}

We reduce the number of layers in the models to reduce evaluation cost.
The estimated throughput and latency of the original model is reported.
The number of layers is strongly proportional to the overall latency, and inversely proportional to the throughput.
So, throughput and latency of the real original model can be directly calculated.
Fairness of comparing systems is not lost by using models of reduced number of layers,
because little chance is found across layers, and no system have done optimizations at the layer dimension.
This is justified by \Cref{fig:layerlat}, showing the latency of Opt-175b model using different number of layers, i.e., the number of transformer blocks.
When keeping other settings unchanged, they are almost linearly related.

To measure the token generation throughput, the models are used to generate a sequence over a short prompt.
The total length of the generated sequences is $1024$ for both models.

The models run over fp16 data format, without any quantization or pruning technique.
Therefore, output of different systems should identical except for floating point errors.

\paragraph{Hardware}
We use a NVIDIA A10 GPU with 24 GB device memory as the \gpuworker{}.
The node has 256 GB host memory as swap space of vLLM.
Up to $4$ additional nodes with dual sockets of AMD Epyc CPUs are used as the \cpuworkers{} of \sys{}.
The cluster is connected via Infiniband network.

As no existing system exploits out-of-chassis CPUs, all the baseline systems run on the GPU node only.
While it may not look fair because \sys{} introduces extra hardware,
existing approaches can only use the CPU nodes as a stand-alone CPU worker for the text generation task,
contributing less than $1\%$ to the total throughput beside the GPU.

\paragraph{Baselines} \textbf{vLLM}~\cite{vllm} uses paged attention technique to manage the \kvcache{}, and efficiently swapping its parts to host memory.
As vLLM is reported to totally outperform \textbf{Orca}~\cite{orca}, we omit Orca in our experiments.

\newcommand{\bllink}[1]{\footnote{\url{#1}}}

\textbf{TensorRT-LLM}\bllink{https://github.com/NVIDIA/TensorRT-LLM} is the newest generation of \textbf{FasterTransformer}\bllink{https://github.com/NVIDIA/FasterTransformer}.
Both systems are developed by NVIDIA with state-of-the-art performance optimizations that can best utilize the GPUs with hardware-specific tuning.

\textbf{FastLLM} \bllink{https://github.com/ztxz16/fastllm} is an accelerated LLM serving system crafted by experts.
It adopts a pure C++ implementation that targets on fast deployment, low latency, and high throughput with various accelerators.

A \textbf{Vanilla}\bllink{https://github.com/facebookresearch/llama} implementation of Llama~\cite{llama,llama2} is released with the model.
The pure PyTorch~\cite{pytorch} implementation and its derived versions are widely used in both academia and industry.
It includes a simple \kvcache{} implementation on GPUs, and achieves competitive throughput thanks to the optimized PyTorch library.

\subsection{Maximum Throughput}

\begin{figure}[ht]
    \centering
    \includegraphics[scale=.5]{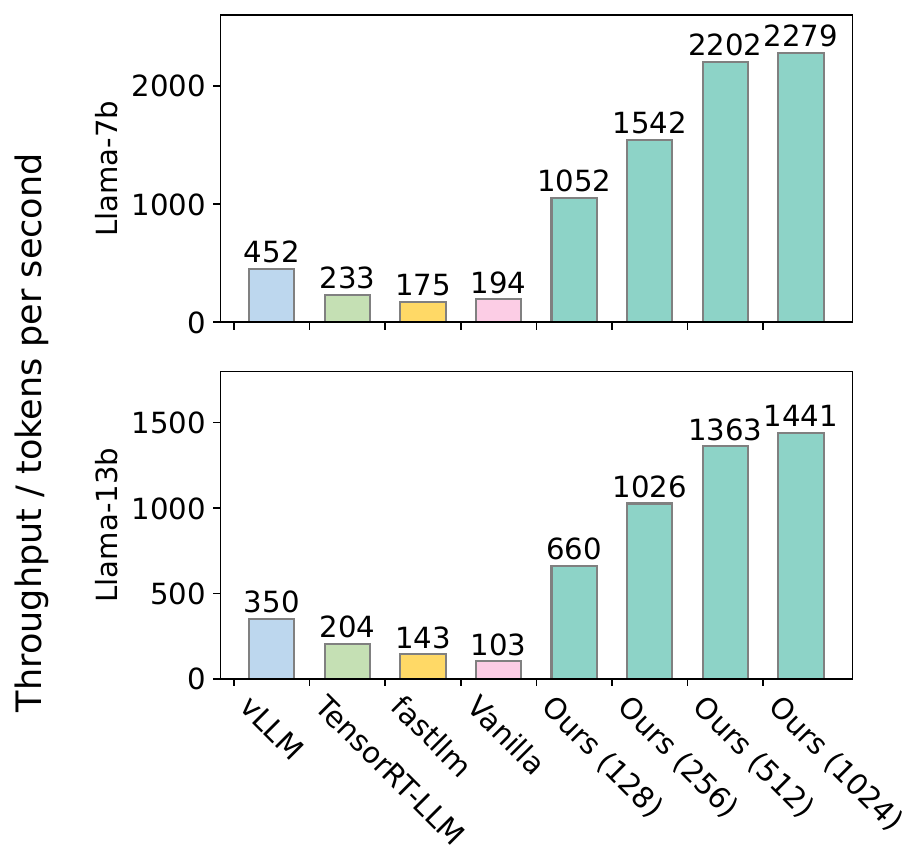}
    \caption{Token generating throughput}
    \label{fig:seqtpt}
\end{figure}

\Cref{fig:seqtpt} shows the measured throughput of all the systems.
The number in brackets after \textit{ours} indicates the batch size of \sys{}.
The possible batch size is enormous in our system, because the distributed host memory is large enough for thousands of sequences.
Increasing the batch size can increase the utilization of GPUs, and thus the overall throughput.
However, as there may be constraint on the latency, the batch size should be set properly.
Also, we observe that the performance gain of increasing batch size gets less when the batch size is large enough.
When the batch size increases by $8\times$ from $128$ to $1024$, we only get $2\times$ throughput.

Comparing with the most powerful baseline, vLLM, in the generation task of the 7b model,
\sys{} achieves a maximum throughput of more than 2k tokens per second, $4\times$ more than vLLM, and $8.7\times$ more than TensorRT-LLM.
For the 13b model, our maximum throughput is $4.12\times$ the throughput of vLLM.
Even when reducing the batch size to 128 for lower latency, we achieve $2.32\times / 1.88\times$ the throughput of vLLM.

When running vLLM, we observe that it can achieve the batch size of $1024$ in the beginning, because the sequences are short, and the \kvcache{} can be all stored in GPU memory.
However, as the sequences get longer, it finds less batching opportunity, and can only use a similar batch size with other GPU-only systems.
TensorRT-LLM performs better than fastllm and vanilla because of its more efficient CUDA kernels.
However, the maximum possible batch size of these systems is barely more than $16$, limited by the GPU memory.
Thus, they have much lower throughput than ours.
The average throughput of our system is $6.71\times$ and $6.04\times$ the average throughput of all baseline systems, respectively.

\subsection{Token Generating Latency}

\begin{figure}[ht]
    \centering
    \includegraphics[scale=.5]{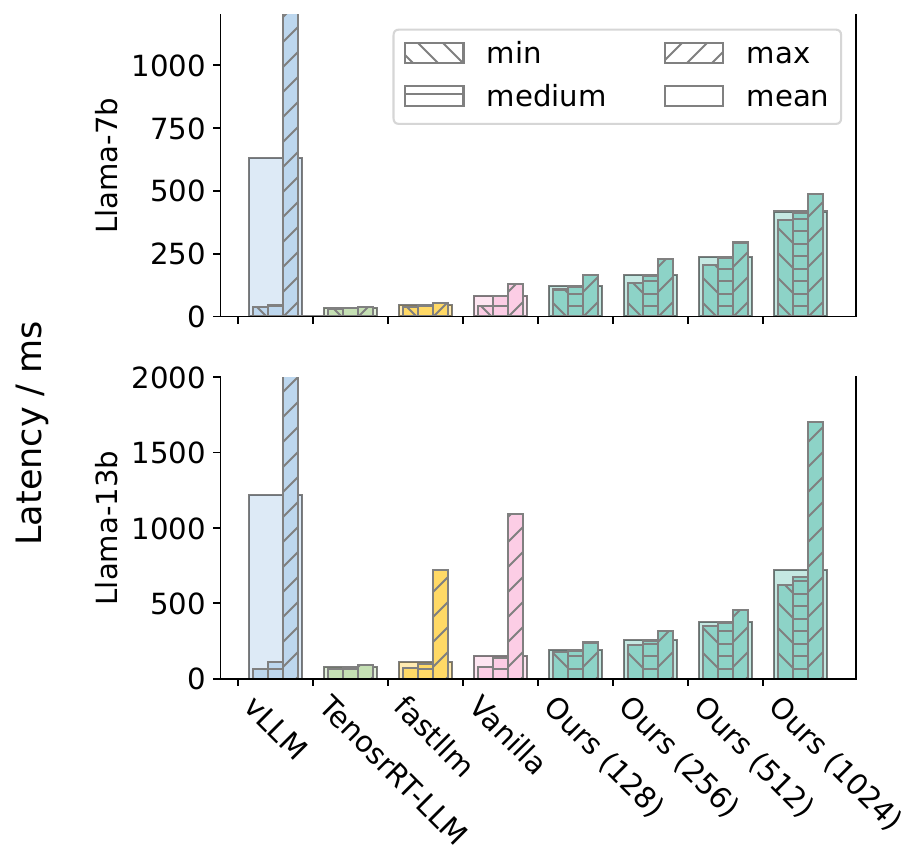}
    \caption{Token generating latency}
    \label{fig:seqlat}
\end{figure}

\Cref{fig:seqlat} shows the measured latency to generate a new token by all the systems.
The wide bar indicates the average latency between generating two adjacent token, and the three narrow bars show $P=0.01/0.5/0.99$ latency, respectively.

When we maximize our batch size to target on highest throughput, the latency is about $3.5\times$ the latency when using $8\times$ smaller batch size.
This also implies the GPU utilization improvement of increased batch size.

TensorRT-LLM achieves the minimum average latency of $34.2$ ms and $77.0$ ms per token, respectively in the two models.
Using a batch size of $128$, the average latency of \sys{} is $120.8$ ms and $191.6$ ms.
With at most $2.5\times$ larger latency of the 7b model, we have $4.5\times$ throughput.
Potentially, given $4\times$ GPUs, we are able to retain the throughput improvement while reducing the per-token average latency to the same level as TensorRT-LLM.

The latency of vLLM is as low as other systems when generating most tokens, because it uses a similar small batch size in most steps.
However, the average latency of vLLM is higher than all setups of \sys{}.
This is because a few steps that swaps the \kvcache{} between host and GPU memory are significantly slow, a key bottleneck of all systems that offloads the \kvcache{}.

\section{Performance Analysis}
\label{sec:ablation}

\subsection{Coping with Heterogeneity}

\begin{figure}[h]
    \centering
    \includegraphics[scale=.5]{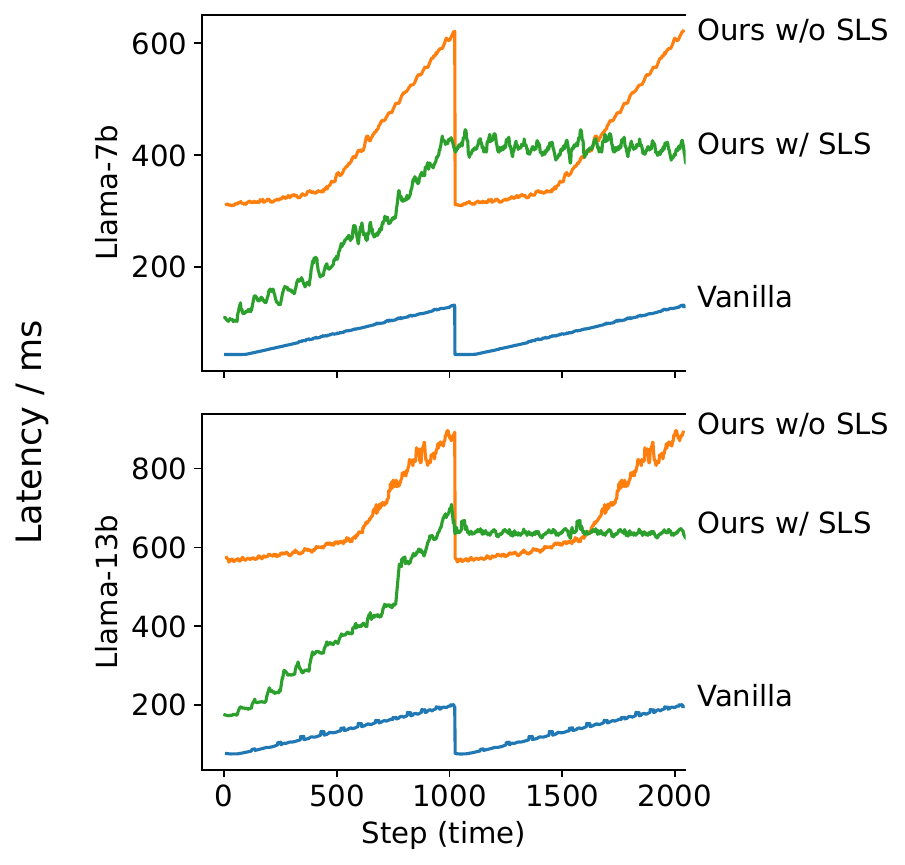}
    \caption{Latency at each step}
    \label{fig:latcurve}
\end{figure}

The curve of per-step latency in \Cref{fig:latcurve} clearly shows the difference between \sys{} with or without the sequence-level load-stabilizing schedule (SLS).

As a baseline, the vanilla implementation runs both \rpart{} and \spart{} on the same GPU.
So, its latency grows linearly with the workload of \rpart{}, which is proportional to the length of the current sequences.

Without the load-stabilizing schedule, the latency first grows slightly.
While most part of \spart{} and \rpart{} are overlapped by the basic two-stage pipeline, the starting and ending overhead is exposed, and leads to the gentle increase of overall latency.
Then, after the length of generated sequences exceeds a certain point, the latency grows sharply with time, because the latency becomes dominated by the increasing latency of \rpart{}, 
and the GPU gets underutilized.

After a cold start process of latency and low throughput due to smaller batch size, 
the sequence-level load-stabilizing schedule provides a stable latency at $66\% - 70\%$ the maximum latency without it.
The sustainable throughput is increased by $8\% - 11\%$ by the technique.
Overhead of the pipeline stops the system from achieving the ideal benefit of $50\%$ maximum latency reduction and $20\%$ throughput improvement indicated by \Cref{fig:slidle},

The smaller improvement on the 7b model compared with the 13b model is also caused by overload of the \cpuworkers{}.
Feature dimension $h$ of the 13b model is larger than the 7b model.
The workload of fully-connected layers in \spart{} is proportional to $O(h^2)$, while it is $O(h)$ in \rpart{}.
Therefore, it is expected that \rpart{} have more workload than \spart{} in smaller models.

\begin{figure}[h]
    \centering
    \includegraphics[scale=.5]{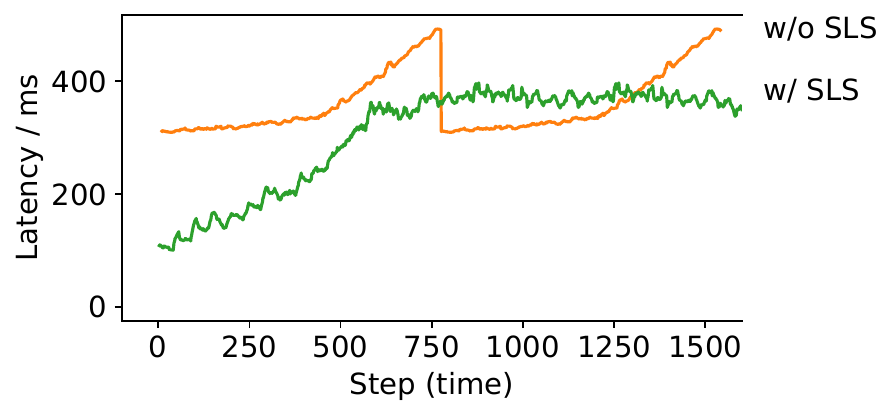}
    \caption{Latency at each step for Llama-7b model with reduced sequence length}
    \label{fig:latcurveo}
\end{figure}

As indicated by \Cref{equ:cpucon}, the sequence length $S$ is proportional to the required number of \cpuworkers{}.
To justify the performance model, as shown in \Cref{fig:latcurveo}, we reduce the length of generated sequences to $768$.
The latency of \spart{} gets closer to the sustainable latency of the sequence-level load-stabilizing schedule, indicating more balanced workloads between the \gpuworker{} and the \cpuworkers{}.
The throughput improvement increases from $8\%$ to $13\%$.

\subsection{Scalability}

Employing more CPUs is a basic requirement of \sys{} for certain workload.
So, scalability of the \cpuworkers{} is important to the overall efficiency.
We use a fixed workload, generating tokens after $1024$ sequences of length $1024$ or $128$.
Each worker is bound to a socket.
We evaluate the scalability of \sys{} on up to $8$ sockets on $4$ nodes.

\begin{figure}[ht]
    \centering
    \includegraphics[scale=.5]{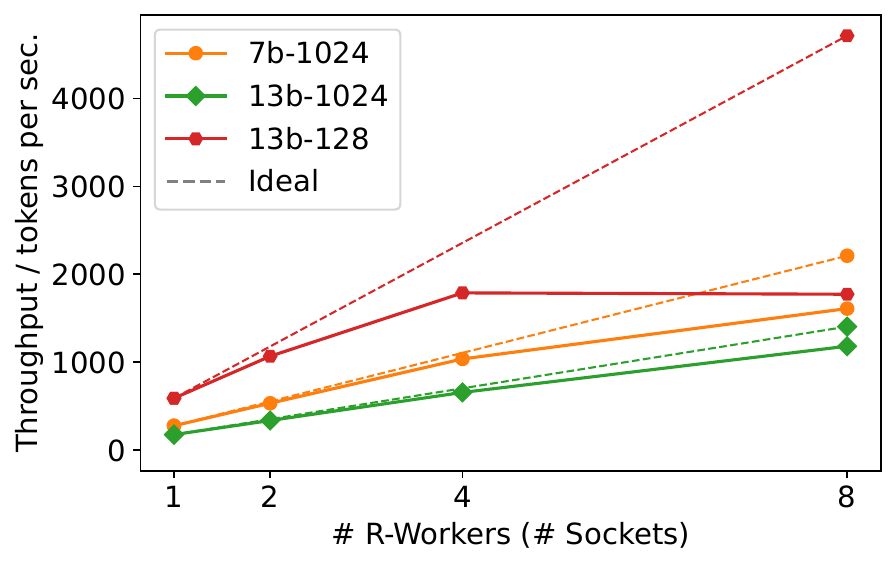}
    \caption{Strong scalability of \sys{}}
    \label{fig:scalability}
\end{figure}

\Cref{fig:scalability} shows the strong scalability experiment results of \sys{} over the 7b and 13b model. 
When the length of sequences is $1024$, \sys{} achieves $72.8\%$ and $84.1\%$ efficiency when scaling up from $1$ socket to $8$ sockets, on the 7b and 13b models, respectively.
As the total latency is smaller in the 7b model, overhead of the pipeline is more significant, leading to lower efficiency with $8$ sockets.
When sequence is as short as $128$, the efficiency is $37.6\%$ for the 13b model.
Using $8$ sockets achieves even lower throughput than using $4$ sockets with $75.9\%$ efficiency.
This is implied by our performance model.
Shorter sequences require less \cpuworkers{}.
Employing more \cpuworkers{} does not increase the performance when the \gpuworker{} is the bottleneck.

\begin{figure}[ht]
    \centering
    \includegraphics[scale=.5]{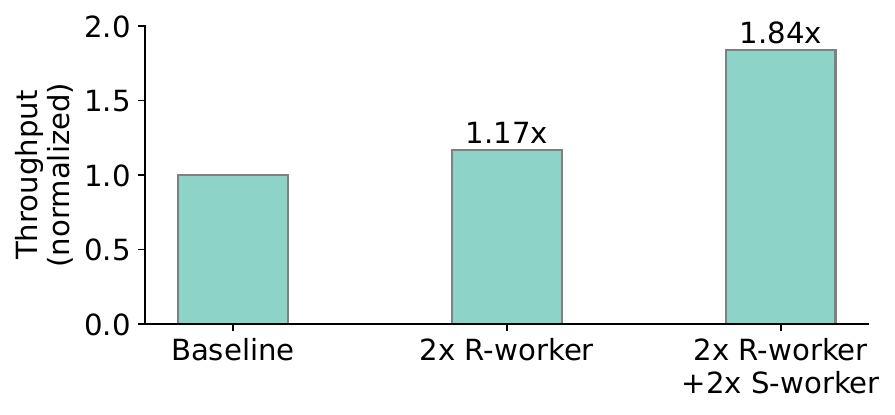}
    \caption{Using more workers in \sys{}}
    \label{fig:mp}
\end{figure}

We show the ability of scaling up \sys{} by using more \gpuworkers{} with model parallelism.
In this case, we use the Opt-175b model which requires less \cpuworkers{} to work with a \gpuworker{},
because the Opt model is larger, so less \cpuworkers{} are needed, as \Cref{equ:cpucon} suggests.
As a baseline setting, we use one A10 GPU with two Epyc CPUs in a node.
Both hardware are well utilized, while the \cpuworkers{} are slightly overloaded.
\Cref{fig:mp} shows the results of introducing more hardware.
When only using $2\times$ CPUs as \cpuworkers{}, the overall throughput is only slightly increased.
For the bar on the right, we double the number of both \cpuworkers{} and \gpuworkers{}.
The two \gpuworkers{} work in model parallelism by partitioning all the parameter and activation tensors. 
\sys{} achieves $1.84\times$ throughput using double hardware.

\subsection{Latency Break-down}

\begin{figure}[h]
    \centering
    \includegraphics[scale=.5]{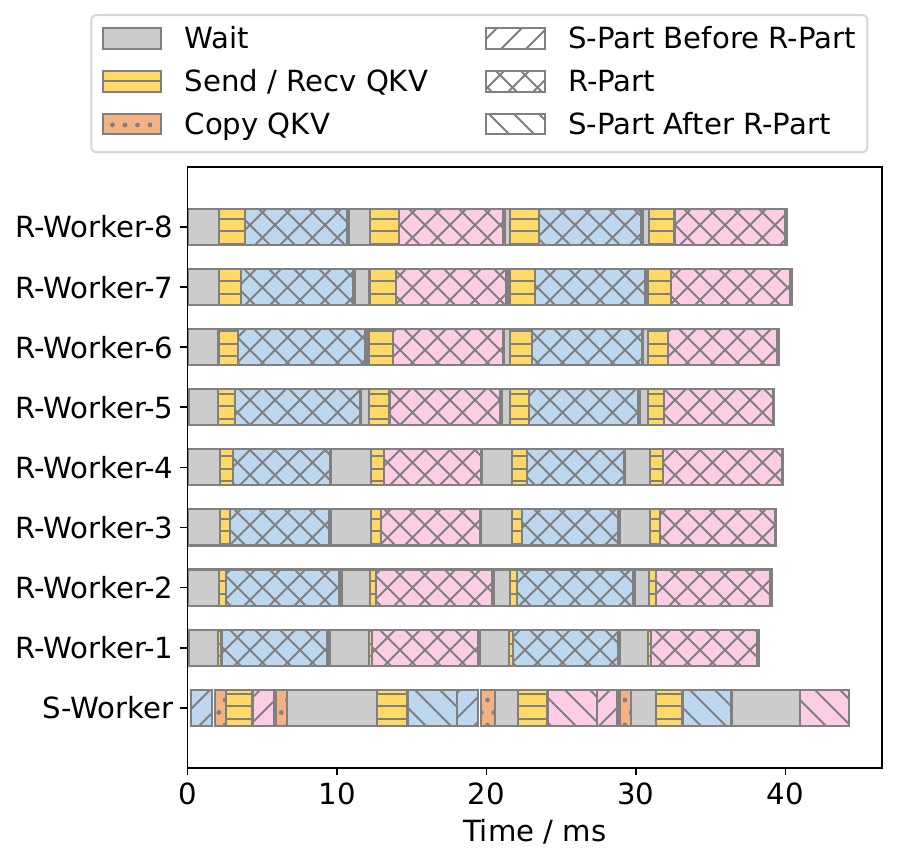}
    \caption{Latency of two layers in a 13b model}
    \label{fig:breakdown}
\end{figure}

To see the detailed utilization of different workers, we trace different operations of the workers, as shown in \Cref{fig:breakdown}.
The \cpuworkers{} are busy with computation in more than $75\%$ of the time.
The performance variance across nodes makes some of the workers wait for others.

Copying the QKV data from GPU to CPU takes $3$ ms, during the iteration of $43$ ms to generate a new token.
Sending QKV across the network takes another $7.4$ ms.
In total, the distributed design of \sys{} introduces about $25\%$ overhead to transmit the feature vectors.
Notably, we change the asynchronous communication to synchronous mode for profiling.
In production, the asynchronous communication can overlap part of the communication overhead.

The \gpuworker{} is actually working in less than $50\%$ of time in the profiled case, due to overloading and performance variance of the \cpuworkers{}.
However,  the overall throughput is still competitive,
as the efficiency to compute \spart{} is significantly increased because of the much larger batch size.
\section{Related Works and Discussion}
\label{sec:related}

\paragraph{Optimizing the Attention Operator}
For training LLMs, FlashAttention~\cite{flashattn,flashattn2} is a widely-used optimized fused attention operator.
It achieves performance improvement by eliminating the need to store the memory-consuming intermediate $A$ matrix.
The idea is ported to the token generation scenario by FlashDecoding~\cite{flashdecoding}.
However, as $A$ is a vector of much smaller size in decoding, it has less impact than that in training.
These techniques can be ported to CPUs and accelerate our computation.

The idea of window attention~\cite{longformer}, which is further extended in StreamingLLM~\cite{streamingllm}, is a variation of the original attention algorithm that reduces the number of tokens to interact with for each new token.
Our system benefits from these techniques in the same way as quantization~\cite{awq,gptq,quant} and pruning~\cite{powerinfer,fastpruning}.
They reduce the workload of \rpart{}, while the user has to be are of the potential change of the model quality.

Speculative token generation~\cite{specinfer} is currently the only approach that essentially increases the efficiency of attention operation,
which depends on accurate prediction of tokens.

\vspace{-1em}
\paragraph{Distributed and Heterogeneous LLM Serving}
Typical model partitioning systems~\cite{flexflow,alpa,alpaserve} can barely handle the token generation case where hardware and workload of the model are both heterogeneous.
Beside offloading \kvcache{} to host memory~\cite{vllm,flexgen,swapadvisor,deepum}, peer GPUs~\cite{lin2024infinitellm} may also be the place to offload, despite the expense.
In fact, FPGAs~\cite{chen2023understanding,zeng2024flightllm} may be a better choice to store and process \kvcache{}.

The idea of this paper can be generalized as using heterogeneous hardware for the different parts of LLMs for better efficiency of the whole system.
Beside CPU, possible selection of hardware for the memory-intensive part includes cheaper GPUs, FPGAs, and domain-specific chips.
A memory pool directly connected to the GPU by CXL~\cite{cxl} would be a even more reasonable option for the \kvcache{}.
Among the various possibilities, our approach that unleashes the power of CPUs for \rpart{} is an immediately feasible approach that uses existing hardware with high affordability.
\section{Conclusion}
\label{sec:conclu}

In this paper, we propose \sys{}, a system that achieves high throughput of generating tokens with LLMs using affordable GPU resources.
Different from typical solutions that fully use GPUs for computation,
we decompose the model into two parts, and move both storage and computation of the memory-bound part to distributed out-of-chassis CPUs, utilizing their aggregated compute power.
Performance challenges brought by heterogeneity in both temporally varying workload and hardware are addressed by a sequence-level load-stabilizing schedule and a performance model.
Finally, as the GPU is better utilized thanks to greatly enlarged batch size, and the overall throughput is competitive.

\bibliographystyle{plain}
\bibliography{ref}

\end{document}